\documentclass[fleqn,usenatbib]{mnras}

\usepackage[normalem]{ulem}

\usepackage{newtxtext,newtxmath}

\usepackage[T1]{fontenc}

\DeclareRobustCommand{\VAN}[3]{#2}
\let\VANthebibliography\thebibliography
\def\thebibliography{\DeclareRobustCommand{\VAN}[3]{##3}\VANthebibliography}

\usepackage{graphicx}	
\usepackage{amsmath}	

\title[Improving pulsar search efficiency with AI]{Improving pulsar search efficiency in next-generation pulsar surveys with artificial intelligence}

\author[Q.~Y.~Fu et al.]{
Qiuyang~Fu$^{\href{https://orcid.org/0000-0003-0307-5633}{\includegraphics[scale=0.08]{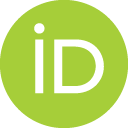}}}$,$^{1,2}$\thanks{E-mail:fuqy@bao.ac.cn}
Mengyao~Xue$^{\href{https://orcid.org/0000-0001-8018-1830}{\includegraphics[scale=0.08]{Figure/ORCIDiD.png}}}$,$^{1}$\thanks{mengyaoxue@nao.cas.cn}
Weiwei~Zhu$^{\href{https://orcid.org/0000-0001-5105-4058}{\includegraphics[scale=0.08]{Figure/ORCIDiD.png}}}$,$^{1,3}$
N.~D.~R.~Bhat$^{\href{https://orcid.org/0000-0002-8383-5059}{\includegraphics[scale=0.08]{Figure/ORCIDiD.png}}}$,$^{4}$
Kaichao~Wu$^{\href{https://orcid.org/0009-0007-9367-730X}{\includegraphics[scale=0.08]{Figure/ORCIDiD.png}}}$,$^{5}$
\newauthor
~Zihan~Zhang$^{\href{https://orcid.org/0009-0000-0676-3277}{\includegraphics[scale=0.08]{Figure/ORCIDiD.png}}}$,$^{5,6}$
B.~W.~Meyers$^{\href{https://orcid.org/0000-0001-8845-1225}{\includegraphics[scale=0.08]{Figure/ORCIDiD.png}}}$,$^{7,4}$
Chia~Min~Tan$^{\href{https://orcid.org/0000-0001-7509-0117}{\includegraphics[scale=0.08]{Figure/ORCIDiD.png}}}$,$^{4}$
Youling~Yue$^{\href{https://orcid.org/0000-0003-4415-2148}{\includegraphics[scale=0.08]{Figure/ORCIDiD.png}}}$,$^{1}$
Jiarui~Niu$^{\href{https://orcid.org/0000-0001-8065-4191}{\includegraphics[scale=0.08]{Figure/ORCIDiD.png}}}$,$^{1}$ 
\newauthor
~Lingqi~Meng$^{\href{https://orcid.org/0000-0002-2885-568X}{\includegraphics[scale=0.08]{Figure/ORCIDiD.png}}}$,$^{1,2}$ 
Ziwei~Wu$^{\href{https://orcid.org/0000-0002-1381-7859}{\includegraphics[scale=0.08]{Figure/ORCIDiD.png}}}$,$^{1}$
Ziyao~Fang$^{\href{https://orcid.org/0009-0008-3048-2641}{\includegraphics[scale=0.08]{Figure/ORCIDiD.png}}}$,$^{1}$
Yukai~Zhou$^{\href{https://orcid.org/0009-0005-4137-7693}{\includegraphics[scale=0.08]{Figure/ORCIDiD.png}}}$,$^{8}$
Jiawei~Jin$^{\href{https://orcid.org/0009-0004-2538-2321}{\includegraphics[scale=0.08]{Figure/ORCIDiD.png}}}$$^{1,2}$
\\
$^{1}$National Astronomical Observatories, Chinese Academy of Sciences, 20A Datun Road, Chaoyang District, Beijing 100101, China \\ 
$^{2}$School of Astronomy and Space Science, University of Chinese Academy of Sciences, Beijing, 100049, China \\ 
$^{3}$Institute for Frontiers in Astronomy and Astrophysics, Beijing Normal University, Beijing 102206, China \\
$^{4}$International Centre for Radio Astronomy Research (ICRAR), Curtin University, Bentley, WA 6102, Australia \\ 
$^{5}$Computer Network Information Center, Chinese Academy of Sciences,\\ \hspace{3pt}CAS Informatization Plaza No.2 Dong Sheng Nan Lu, Haidian District, Beijing 100083, China \\ 
$^{6}$School of Computer Science and Technology, University of Chinese Academy of Sciences, Beijing, 100049, China \\ 
$^{7}$Australian SKA Regional Centre (AusSRC), Curtin University, Bentley, WA 6102, Australia \\ 
$^{8}$Department of Astronomy, School of Physics, Peking University, Beijing 100871, China \\ 
}

\date{Accepted XXX. Received YYY; in original form ZZZ}

\pubyear{\the\year{}}

\begin{document}
\label{firstpage}
\pagerange{\pageref{firstpage}--\pageref{lastpage}}
\maketitle

\begin{abstract}
Pulsar searching with next-generation radio telescopes requires efficiently sifting through millions of candidates generated by search pipelines to identify the most promising ones. This challenge has motivated the utilization of Artificial Intelligence (AI)-based tools. In this work, we explore an optimized pulsar search pipeline that utilizes deep learning to sift ``snapshot'' candidates generated by folding de-dispersed time series data. This approach significantly accelerates the search process by reducing the time spent on the folding step. We also developed a script to generate simulated pulsars for benchmarking and model fine-tuning. The benchmark analysis used the NGC 5904 globular cluster data and simulated pulsar data, showing that our pipeline reduces candidate folding time by a factor of $\sim$10 and achieves 100\% recall by recovering all known detectable pulsars in the restricted parameter space. We also tested the speed-up using data of known pulsars from a single observation in the Southern-sky MWA Rapid Two-metre (SMART) survey, achieving a conservatively estimated speed-up factor of 60 in the folding step over a large parameter space. We tested the model's ability to classify pulsar candidates using real data collected from the FAST, GBT, MWA, Arecibo, and Parkes, demonstrating that our method can be generalized to different telescopes. The results show that the optimized pipeline identifies pulsars with an accuracy of 0.983 and a recall of 0.9844 on the real dataset. This approach can be used to improve the processing efficiency for the SMART and is also relevant for future SKA pulsar surveys. 
\end{abstract}




\section{Introduction}
Pulsars are rapidly spinning neutron stars with strong gravitational and magnetic fields. The discovery of the first pulsar, PSR 1919+21, marked the advent of pulsar astronomy \citep{1968Natur.217..709H}. The first binary pulsar system, PSR B1913+16, provided the first evidence of gravitational waves \citep{hulse1975discovery}. The discovery of radio pulsars relies on telescopes to capture signals originating from beyond our solar system, which have traversed the interstellar medium (ISM). The radio waves interact with free electrons in the ISM during propagation, leading to dispersion, whereby low-frequency signals arrive later than high-frequency signals. This effect is quantified by the dispersion measure (DM), a value directly proportional to the integrated column density of free electrons between the pulsar and observer \citep{lorimer2005handbook}. Additionally, pulsar observations need to account for other propagation effects of the ISM, such as scattering (i.e., pulse broadening) \citep{jing2025fast} and Faraday rotation. These effects are significant at low frequencies ($\lesssim$300 MHz), near the Galactic center, or along the Galactic plane \citep{rickett1990radio}. 

The vast majority of pulsars, particularly millisecond pulsars, are only moderately luminous and as a result single pulses are difficult to observe directly \citep{taylor1991millisecond}. To enhance the signal-to-noise ratio (SNR), a common technique is to fold the observational data according to the pulsar's rotational period. By aligning and summing multiple pulses, this folding process amplifies the periodic signal above the noise background. This approach is essential for the discovery of a substantial fraction of pulsars. 

The number of candidates produced by modern pulsar surveys exceeds the capacity for manual inspection. In 1977, the second Molonglo Survey identified around 2,500 possible candidates across the sky south of declination $+20^\circ$ \citep{manchester1978second}. Three decades later, the Northern part of the High Time Resolution Universe (HTRU) survey found over 80 million candidates across the entire region observed above $+30^\circ$ declination \footnote{\url{https://www.jb.man.ac.uk/pulsar/surveys.html}} \citep{barr2013northern}. Upcoming surveys with the Square Kilometre Array (SKA) will generate even larger candidate sets, making manual inspection intractable. Therefore, it is important to develop methods to assist in filtering candidates. 

Early approaches focused on selecting candidate features to build sorting strategies for filtering millions of candidates \citep{keith2009discovery}. Later, \citet{lee2013peace} utilized statistical methods, using six factors from pulsar candidates to determine their identification scores. Meanwhile, machine learning (ML) is increasingly used to screen pulsar candidates. As early as 2010, \citet{eatough2010selection} applied artificial neural network (ANN) algorithms to score candidates for pulsar identification. Subsequently, \citet{zhu2014searching} proposed a classifier combining a convolutional neural network (CNN), an ANN, and a support vector machine (SVM). Several ML classification models have already been applied in pulsar surveys. For example, an ensemble candidate classifier \citep{tan2018ensemble} was used in the LOFAR Tied-Array All-Sky Survey (LOTAAS) \citep{sanidas2019lofar}, and the first-pass processing in the Southern-sky MWA Rapid Two-metre (SMART) survey \citep{bhat2023southern} adopted modest refinements of the LOTAAS classifier. A model incorporating Semi-supervised generative adversarial network (SGAN) \citep{balakrishnan2021pulsar} has been applied for candidate classification in the HTRU survey with the Parkes telescope (Murriyang). A CoAtNet-based classification model \citep{cai2023pulsar} has been adopted in the Galactic Plane Pulsar Snapshot (GPPS) survey \citep{han2021fast} with the Five-hundred-meter
Aperture Spherical radio Telescope (FAST, \citeauthor{jiang2020fundamental} \citeyear{jiang2020fundamental}).

The conventional pulsar candidate classifiers evaluate each candidate folded from the full data. Full data is time-sampled and divided into multi-frequency channels in "SIGPROC filterbank format" \footnote{\url{https://sigproc.sourceforge.net/}}\citep{lorimer2011sigproc} or "PSRFITS format" \footnote{\url{https://psrchive.sourceforge.net/}}\citep{hotan2004psrchive} files. As the observation duration increases, folding time increasingly dominates the pulsar search runtime. So we fold the de-dispersed time series data across stacked frequency channels to reduce computational cost. Our method generates "snapshot" candidates by folding time-series data according to periods, filtering out non-pulsar-like candidates early in the time domain. Then, we fold the full data to obtain complete time and frequency information for further assessment. Consequently, there is a demand for training new feature classifiers specifically adapted to the time domain pulsar features of the candidates.

This paper demonstrates that our pipeline using artificial intelligence (AI) models can significantly reduce the time spent on the folding step, a critical stage in the pulsar search. While AI is mainly used for classifying pulsar candidates, its application in accelerating data processing enhances overall search efficiency. As pulsar surveys produce amounts of data proportional to observation time, current modern low-frequency array surveys such as the LOTAAS, a one-hour integration pulsar survey covering the entire northern sky \citep{sanidas2019lofar}, and the SMART survey, conducted with the Murchison Widefield Array (MWA, \citeauthor{tingay2013murchison} \citeyear{tingay2013murchison}) to cover the sky south of $30^{\circ}$ in declination with 80-minute observation times \citep{bhat2023southern}, are already facing computational challenges from massive data rates. Future surveys, such as upcoming projects with the SKA, will further amplify these demands, making speed-up of the pulsar search process essential. AI solutions tailored to folding steps can reduce processing burdens. 

In this paper, we present an AI-accelerated pulsar Fast Fourier Transform (FFT) search pipeline that minimizes folding time to speed up pulsar searches. In Section \ref{sec:section2}, we introduce an optimized pipeline for the pulsar search, the utilized model architecture, and the applied preprocessing methods for the candidates. In Section \ref{sec:section3}, we describe the data used in this work, including both real and simulated datasets. In Section \ref{sec:section4}, we evaluate both the speed-up of the optimized pipeline and the classification performance of the model. Finally, in Section \ref{sec:section5}, we summarize the method and outline its potential for future pulsar surveys.

\begin{figure*}	  	 
\centering
\includegraphics[width=16cm]{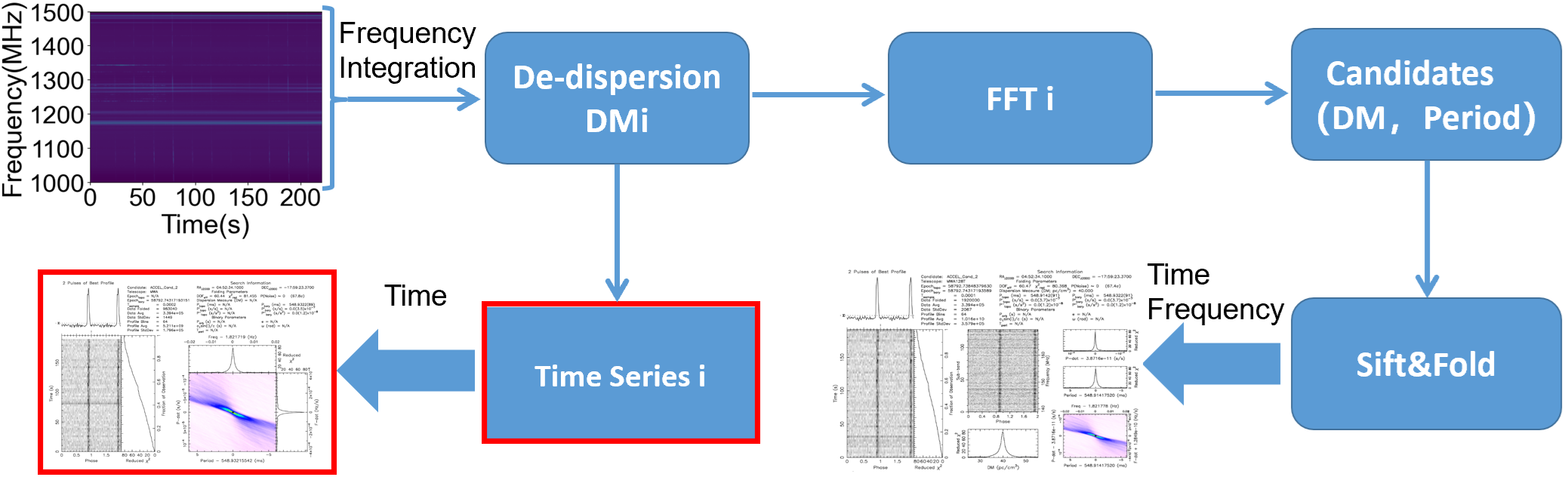}
\caption{
In the FFT-based pulsar search, after trying all possible DM values to remove dispersion delay and applying FFT to get potential periods \citep{cooley1965algorithm}, we obtain a series of candidate DM-period combinations. These candidates can then be folded using the full multi-frequency data or the corresponding de-dispersed time series data.} 
\label{fig:pipeline_01}  
\end{figure*}

\begin{figure*}	  	 
\centering
\includegraphics[width=16cm]{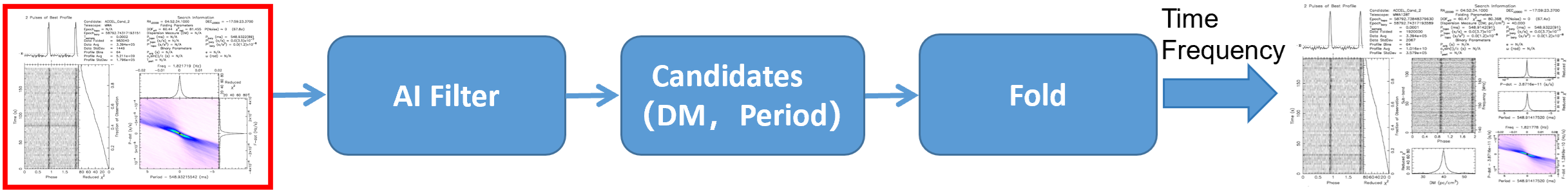}
\caption{We adopt a three-stage approach to reduce folding time. First, we fold the time series data to get initial pulsar candidates. Next, we filter these candidates using an AI classifier based on time-domain features. Finally, we fold the remaining candidates using the full data to confirm their authenticity with additional frequency information.} 
\label{fig:pipeline_02}  
\end{figure*}

\begin{table}
\caption{Pulsar blind search time breakdown with 100 adjacent de-dispersion trials. It presents the number of DM and period combinations generated for 10-, 20-, and 30-minute observation data, as well as the time proportion consumed by each step for these data. The total folding time correlates positively with the number of potential DM and period combinations, and it also increases with longer observations.}
\setlength{\tabcolsep}{4pt}
\renewcommand{\arraystretch}{1.2}
  \begin{tabular}{lccc}
    \hline
    Step & 10-min obs. & 20-min obs. & 30-min obs.  \\
    \hline
    DM–Period combinations & 51 & 81 & 217 \\
    De-dispersion \& FFT   & 1.6\% & 1.1\% & 0.7\% \\
    Accelsearch            & 17.1\% & 14.3\% & 5.9\% \\
    Fold                   & 81.3\% & 84.6\% & 93.4\% \\
    \hline
  \end{tabular}
  \label{tab:search_time}
\end{table}

\section{METHODS} 
\label{sec:section2}

\begin{figure}
\centering
\includegraphics[width=8cm]{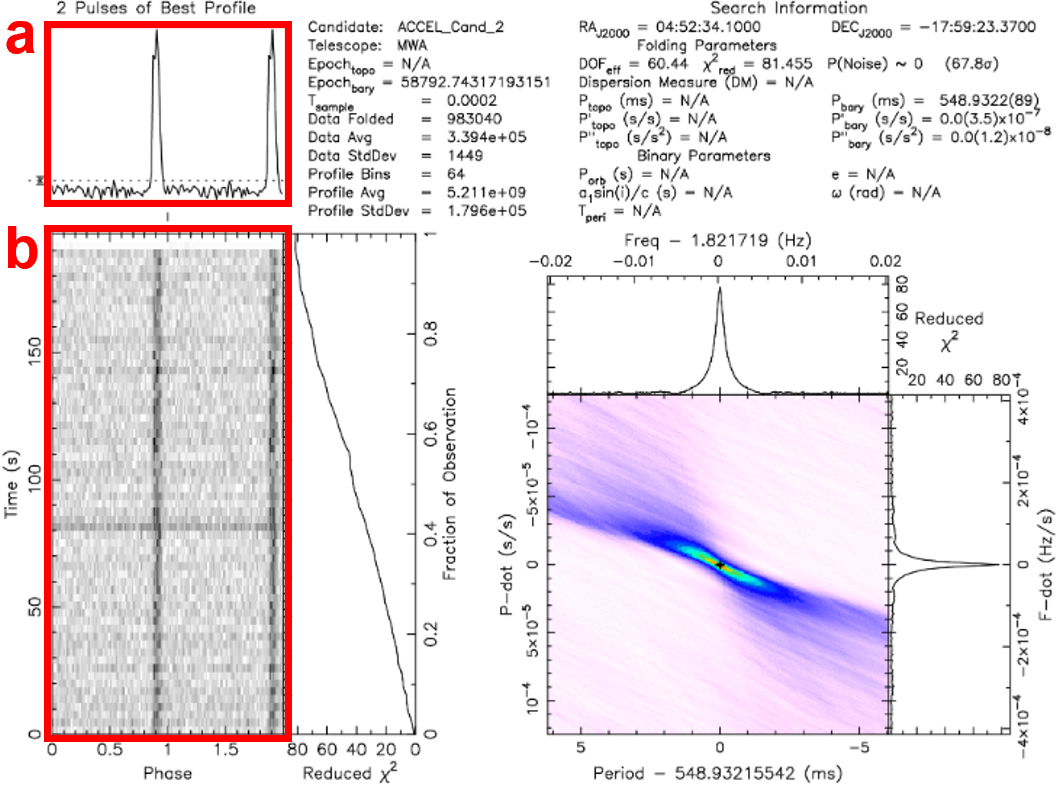}
\caption{The candidate plot exhibits two features: feature "a" corresponds to the average pulse profile obtained by integrating the folded data along the time axis, and feature "b" is a time-phase diagram, where the x-axis shows two phases and the y-axis indicates the integration time.} 
\label{fig:cand}  
\end{figure}



\begin{figure} 	 
\centering
\includegraphics[width=8cm]{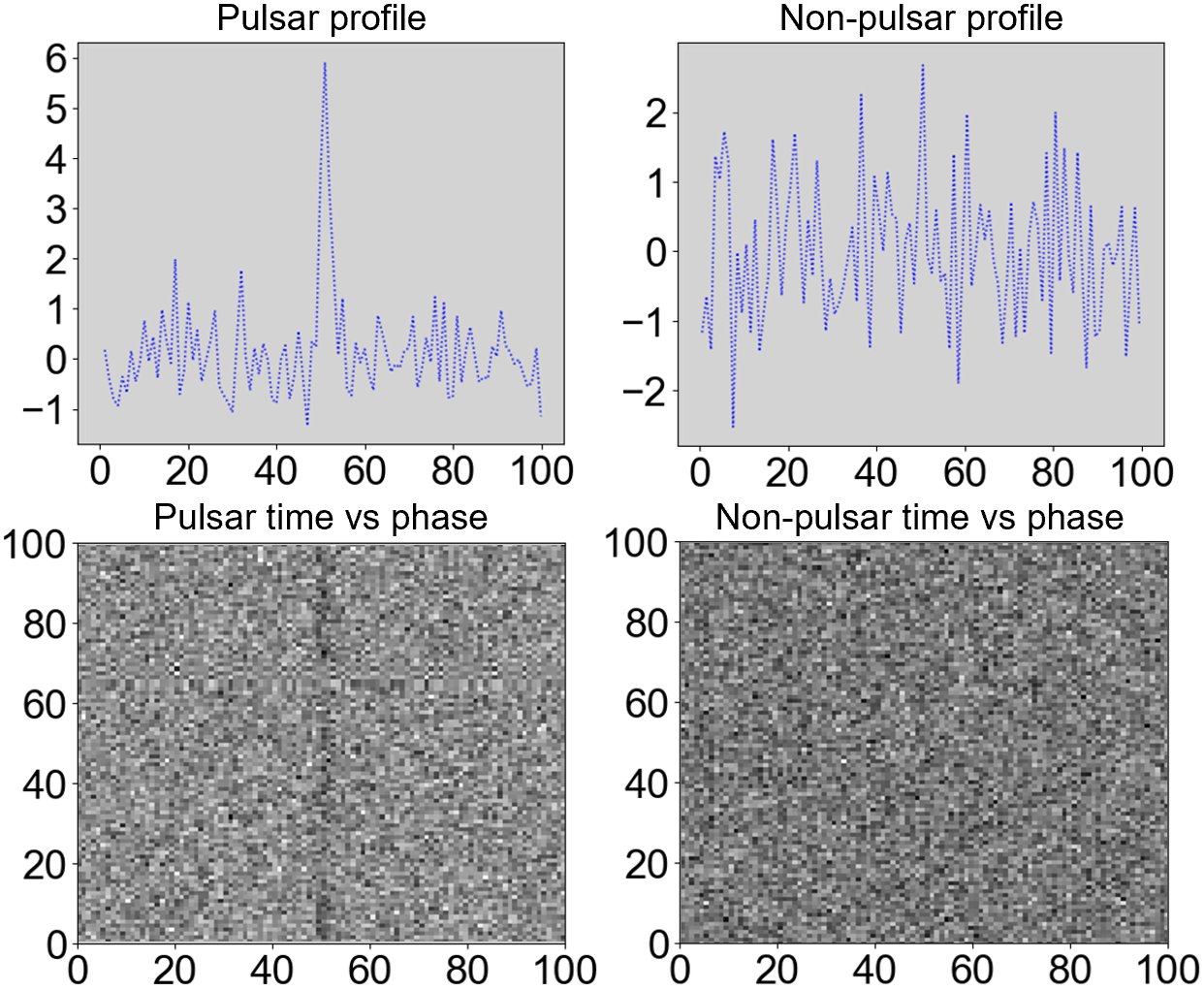}
\caption{Top row: average pulse profiles (feature "a"), where the left panel shows a real pulsar profile characterized by sharp peaks at certain phases due to inherent periodicity and the averaging out of noise, and the right panel shows the profile of a non-pulsar candidate. 
Bottom row: time-phase diagrams (feature "b"), where the left panel illustrates a real pulsar, characterized by periodic signals consistently aligned at the same phase over a time span, whereas the right panel depicts a non-pulsar candidate lacking such phase coherence.}
\label{fig:pulsar_feature}  
\end{figure}


\begin{figure*} 	 
\centering
\includegraphics[width=14cm]
{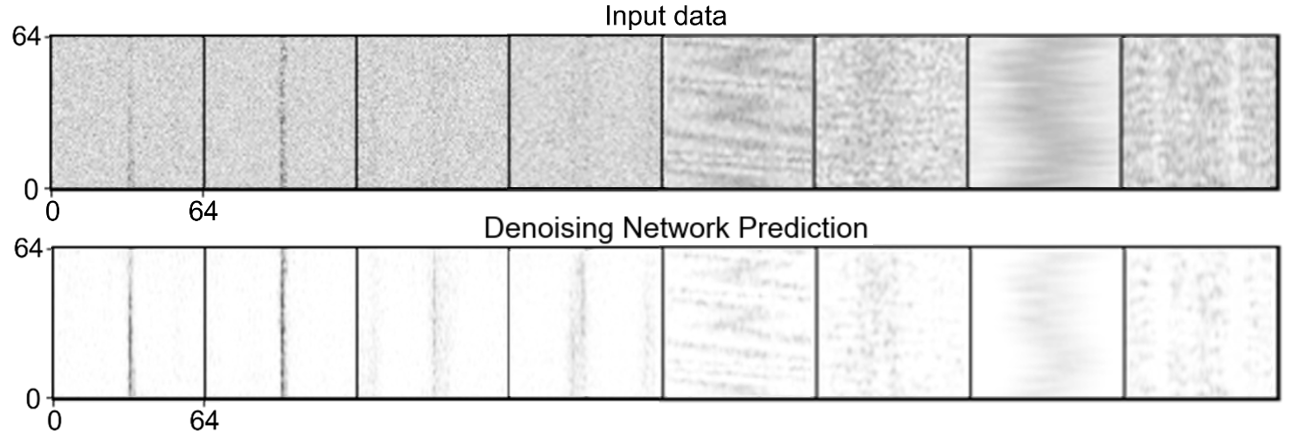}
\caption{We input the time-phase features into a Denoising Autoencoder (DAE). Each instance's x-axis represents the phase bins, while the y-axis corresponds to the time intervals. The example feature arrays are resized to dimensions of 64 by 64. The first four instances are positive samples, and the next four are negative. Using the denoising mechanism of the DAE, it effectively suppresses noise and highlights features.} 
\label{fig:ddpm}  
\end{figure*}

\begin{figure} 	 
\centering
\includegraphics[width=8cm]{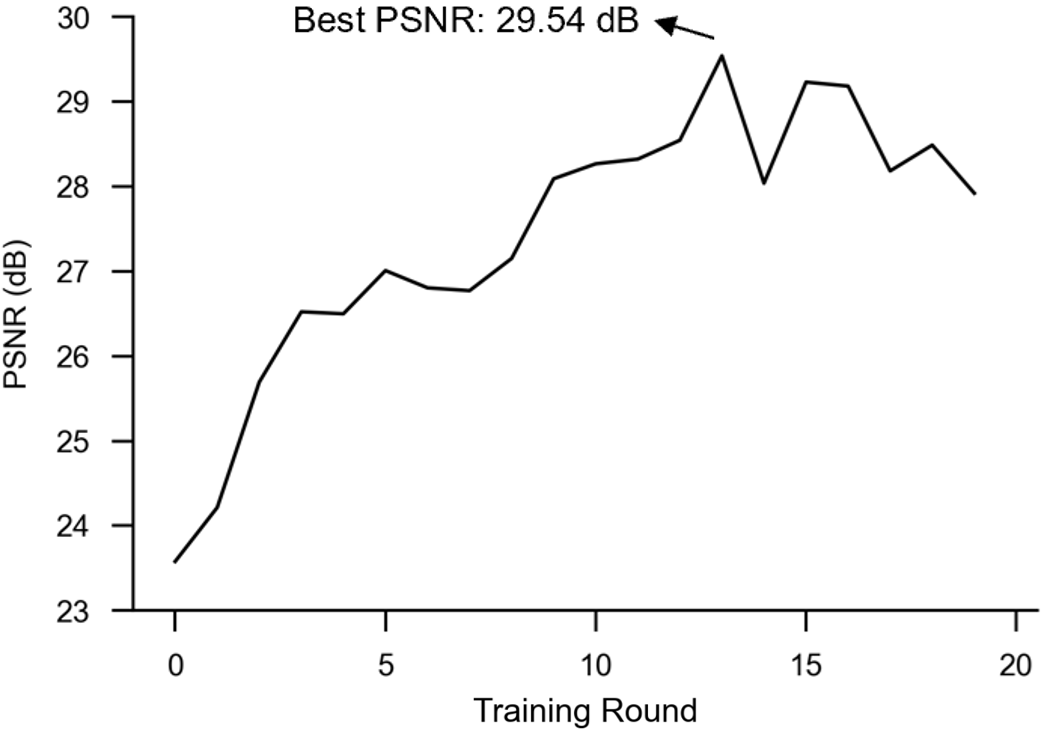}
\caption{The variation of PSNR values calculated from 400 test samples over the training rounds. PSNR quantifies the difference between the original and denoised images, where higher values indicate better denoising and reconstruction quality.} 
\label{fig:psnr}  
\end{figure}

\subsection{Pipeline strategy for speed-up via optimized folding step}
\label{sec:section2.1}

The ATNF Pulsar Catalogue \citep{manchester2005australia} in version 2.6.3 \footnote{\url{https://www.atnf.csiro.au/research/pulsar/psrcat/}} reports over 4,343 pulsars, most of them discovered from pulsar surveys. The PulsaR Exploration and Search TOolkit (PRESTO)\footnote{\url{https://github.com/scottransom/presto}}\citep{ransom2011presto}, has helped to discover over a thousand pulsars. Building on PRESTO's success, we explore an AI-accelerated, FFT-based search pipeline for speeding up the pulsar search process as shown in Figure \ref{fig:pipeline_01}.

The blind search for pulsars based on PRESTO consists of four stages, beginning with the identification and removal of radio frequency interference (RFI) using PRESTO’s ``\textbf{rfifind}'' tool to create a channel mask file. Second, a range of trial DM values was used to generate one-dimensional (1-D) de-dispersed time series for each trial, thereby correcting for unknown dispersion time delays. Each trial generates a separate "\textbf{.dat}" file. Third, the corrected data undergo a Fourier transformation to identify pulsar periods by finding peaks in the power spectrum, and then harmonic summing and peak detection produce candidate parameters for folding. In the final stage, the data are folded for each DM-period combination, generating diagnostic plots to validate potential pulsar candidates. This FFT-based pulsar search involves a two-dimensional (2-D) grid of periods and DM values. Folding with potential combinations enhances the SNR, enabling the detection of weak pulsar signals. While folding full data at specific DM values and periods yields valuable time and frequency information, the computational cost of processing large data with numerous parameter combinations is high.


To quantitatively estimate the proportion of computation time that the folding steps account for in the overall pulsar search process, we processed 30 minutes of FAST observation data (137 GB total). We divided the data into 10-, 20-, and 30-minute segments to analyze the relationship between folding time and observation length, using a single computing node equipped with an Intel Xeon Silver 4314 processor operating at 2.40 GHz. The data contains only total intensity, with a sampling time of $49.152\,\mathrm{\upmu s}$, and 4096 frequency channels. All pulsar search steps run on a single CPU core. We performed 100 adjacent DM trials using a single de-dispersion operation, at nearly the same cost as a single trial. None of the trials contained known pulsars. We applied FFT and acceleration searches using default parameters (zmax = 200, numharm = 8, sigma = 2.0) to each resulting time series, filtered the results with ACCEL\_sift.py, and folded the full data over all DM–period combinations without fine searches. As a result, the Accelsearch and Fold steps dominated the overall computation time. The processing time for each step is detailed in Table~\ref{tab:search_time}. The folding time for potential candidates scales with the number of DM and period combinations, and increases further with longer observation duration. The result shows that during the pulsar blind search process, the folding step constitutes a substantial fraction of the computational time allocation. It presents challenges for large-scale, long-duration pulsar surveys due to the computational cost caused by folding numerous DM and period combinations on full data.

We explore a search strategy to address computational bottlenecks in the folding stage by filtering "snapshot" candidates generated from folding 1-D time series data. This approach substantially decreases the number of full data foldings, thus reducing the usage of computational resources, as shown in Figure \ref{fig:pipeline_02}. Specifically, after de-dispersion with PRESTO, we obtain 1-D de-dispersed time series data by integrating frequencies. 
The time series data occupies a space equal to the full data size divided by the number of frequency channels. We fold this 1-D data at the corresponding DM values using the candidate period. The folded candidate diagnostic plots solely contain time information, as depicted in Figure \ref{fig:cand}. The feature labeled "a" represents the pulse profile, illustrating the radiation characteristics of a pulsar in a 1-D array. Feature "b" is referred to as the time-phase diagram, representing signals that consistently occur at the same phase across an intrinsic time span in a 2-D array. 

Current approaches that integrate ML classifiers into PRESTO-based pipelines such as the LOTAAS search pipeline and the first-pass SMART pipeline. These methods reduce the burden of manually inspecting large numbers of candidates by applying classifiers to folded candidates containing time and frequency information. However, they do not address the computational challenges faced by modern pulsar surveys. Our method first uses candidate parameters to fold 1-D data summed over frequency, then applies a classification model to classify pulsar features, filtering parameters without pulse characteristics in the time domain. This approach allows efficient pre-filtering before folding the full data, reducing the number of folding operations and thereby decreasing the overall processing time.

\begin{figure*}	  	 
\centering
\includegraphics[width=16cm]{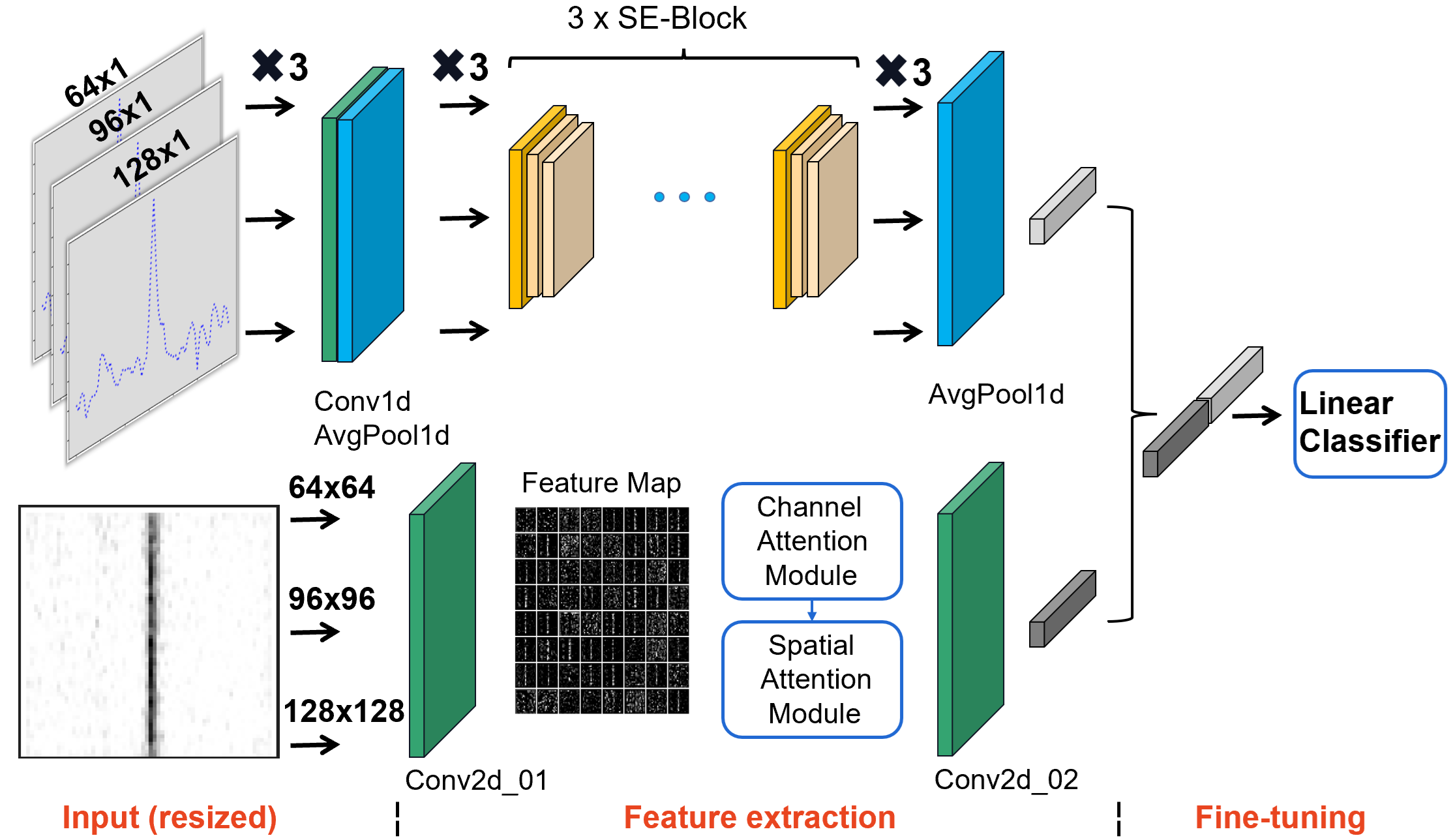}
\caption{An illustration of the model used for sifting "snapshots" candidates. The first step is to resize each candidate feature array to fixed lengths of 64, 96, and 128 bins, respectively, which enables better adaptation to varying candidate data sizes. The resized data is then input into the corresponding feature extraction modules. The extracted features are flattened and fed into a linear classifier to identify pulsar characteristics. When applying the model to candidates from other telescopes outside the training set, a fine-tuning step can be introduced to improve performance.} 
\label{fig:model}  
\end{figure*}

\subsection{Preprocessing of "snapshots" candidates}
\label{sec:section2.2}

Each candidate generated by PRESTO includes an image and a "\textbf{.pfd}" file with candidate information. The features "a" and "b" in the image, as shown in Figure \ref{fig:cand}, are used as identifying characteristics. The array size of these features in the information file is determined by the "-n" parameter used in the folding command, which defines the number of bins (data points) in the profile phase. For instance, the x-axis of the profile in Figure~\ref{fig:pulsar_feature} shows the number of bins within one folded period, which by default corresponds to the number of time samples per period. Consequently, the choice of bin size influences the resulting SNR. However, as neural networks typically require inputs with consistent dimensions, standardizing the data shape becomes necessary. Earlier pulsar classification models resized the input data to a specified size \citep{zhu2014searching,balakrishnan2021pulsar,cai2023pulsar}. It should be noted that this resizing approach can impact the overall performance of the pulsar classifier. Thus, we resize feature arrays to 64, 96, and 128 bins using downsampling and interpolation, creating three distinct samples for a multi-input fusion model. This approach minimizes the impact on classification performance caused by resizing feature arrays to a uniform size. In addition, it increases the width of the network, enabling the model to capture features across multiple scales.

After extracting data from the "\textbf{.pfd}" file, the existence of weak candidates increases the risk of features being overshadowed by noise. We use a Denoising Autoencoder (DAE) \citep{vincent2008extracting} for denoising in the data preprocessing stage, focusing on time-phase features (bottom two panels of Figure \ref{fig:pulsar_feature}). The DAE creates noisy inputs by adding random noise to clean data and is trained to output a reconstruction of the original clean samples. Its objective is to minimize the difference between the reconstructed output and the clean input, making the model capable of denoising and recovering structural features. We implement the DAE with a U-Net \citep{ronneberger2015u}, adapted for image-to-image translation, to remove noise. In this setup, U-Net serves as the network architecture of the DAE, which is effective for denoising because its encoder-decoder architecture uses skip connections to pass high-resolution features from encoder to decoder, enabling the model to suppress noise while preserving fine structural details. We selected 4,000 positive pulsar samples as the original images and adopted a training strategy that adds increasing levels of Gaussian noise with more rounds \citep{ho2020denoising}. This generates noisy positive samples for the model to learn denoising. Before training, we randomly sampled 400 examples for evaluation. We used the peak signal-to-noise ratio (PSNR) between the original and denoised images to assess reconstruction quality \citep{jahne2005digital}. This process involves training the U-Net model on a dataset of noisy original images, using a loss function that penalizes the discrepancy between the denoised output and the original image. See appendix~\ref{sec:advanced} for the definitions of PSNR and the loss function. The U-Net architecture, consisting of contracting and expansive paths, captures context and localizes features precisely, effectively denoising images by preserving structures and suppressing noise artifacts. Table~\ref{tab:basic_unet} summarizes the model we used. After training, the U-Net model effectively transforms noisy inputs into relatively clean ones. This process enhances the quality of time-phase features, as shown in Figure \ref{fig:ddpm}. And the PSNR results of the denoising model are shown in Figure~\ref{fig:psnr}. The results show that the denoising model enhances the weak features of real pulsar samples embedded in heavy noise.

\subsection{AI model for sifting "snapshots" candidates}
\label{sec:section2.3}

We used two features, the pulse profile and the time-phase ("a" and "b" in Figure \ref{fig:cand}), to score the pulsar candidates. As introduced in the previous Section \ref{sec:section2.2}, we preprocess the candidate data by resizing it to various sizes and denoising the 2-D data. We then extract features from the profile using a one-dimensional residual network (1-D ResNet). ResNet is designed to pass information forward more effectively by adding shortcut connections between layers, which prevents the loss of important information when the model becomes deep \citep{he2016deep}. Our model requires sufficient depth to learn the characteristics of the profile features, while avoiding the risk that deeper models extract more generalized patterns and overlook important details. So we use a simplified, customized ResNet model to ensure effective learning of the distinguishing characteristics of the profile features. We further incorporate a Squeeze-and-Excitation attention (SE attention) mechanism, which helps the model focus on important parts of the profile features. It works by adaptively reweighting different convolutional channels, each corresponding to a feature map. Each channel acts as a specialized filter that highlights a specific aspect of profile features, allowing the model to emphasize the most informative representations \citep{hu2018squeeze}. For time-phase features, we use CNNs with the Convolutional Block Attention Module (CBAM), which enables the model to focus on the most informative spatial and channel regions \citep{woo2018cbam}. It focuses on important positions in the time-phase feature map for spatial information and on the most relevant feature channels for channel information. Following feature extraction, a fully connected layer with each input connected to all outputs is used to classify the extracted features. It produces two logits, which are unnormalized scores representing the model’s raw predictions for the binary classes. These logits are converted into a probability distribution through a Softmax activation, ensuring non-negative values and normalization. The model is trained by minimizing the cross-entropy loss, which quantifies the difference between the predicted probabilities and the true labels. For a single sample, the loss is defined as: 
\begin{equation}
L = - \sum_{i=1}^{2} y_i \log(p_i)
\end{equation} 
where $y_i$ is the one-hot encoded label for class $i$ in the binary classification, with $[y_0, y_1] = [1,0]$ for the non-pulsar class and $[y_0, y_1] = [0,1]$ for the pulsar class, and $p_i$ is the predicted probability for class $i$ obtained by the Softmax function. This hybrid model, as shown in Figure \ref{fig:model}, handles multi-size inputs and employs different neural networks for each feature, thereby enhancing feature extraction capability. We implemented the hybrid model with Tinygrad \footnote{\url{https://github.com/tinygrad/tinygrad}}, a lightweight deep learning framework that supports various hardware accelerators. The details of this model are presented in Table~\ref{tab:combined_model}. In the following paragraphs, we introduce each module and provide detailed descriptions of how they are used.

CNNs are deep learning models excelling in computer vision tasks like image classification and object detection \citep{lecun1998gradient, krizhevsky2012imagenet}. They use convolutional layers to extract features and fully connected layers to make predictions. 
They include pooling layers to reduce computation and memory usage, and activation functions like ReLU \citep{nair2010rectified} and sigmoid to introduce nonlinearity. Techniques such as dropout and batch normalization \citep{ioffe2015batch} further enhance performance by preventing overfitting and improving training stability. Inspired by GoogLeNet's multi-scale approach \citep{szegedy2015going}, our architecture replaces parallel multi-kernel convolutions and pooling with multi-sized input resizing. This enables simultaneous capture of multi-scale features within a single layer, broadening network width to enhance model performance.

ResNet is a deep neural network architecture that uses residual connections to train deeper networks effectively and address vanishing gradients \citep{he2016deep}. Our bottleneck residual block, consisting of three convolutional layers and a shortcut connection, facilitates this process. We use a 1-D ResNet with SE attention to classify pulse profile features. The SE attention mechanism helps the network focus on the most relevant features, improving classification accuracy. The profile features are resized to dimensions of 64, 96, and 128 bins. The network processes these resized features to extract convolutional features, which are flattened, concatenated, and input into a linear classifier for pulsar classification. The ResNet architecture allows capturing discriminative features from diagnostic plots at multiple levels, aiming for better classification performance.

CBAM enhances CNN representational power by integrating channel and spatial attention mechanisms \citep{woo2018cbam}. The channel attention recalibrates feature responses to highlight informative channels, while spatial attention adjusts responses based on spatial relationships to focus on important regions. We preprocess candidate time-phase features using a DAE for denoising, then resize the data to 64, 96, and 128 bins. Each resized array is then processed through convolutional layers to extract features. The attention module applies channel attention before spatial attention, allowing the spatial mechanism to act on more informative channel features. These attention mechanisms are applied to the feature maps to compute weights that emphasize important regions within them. Then these features are flattened, concatenated, and input into a linear classifier.

Fine-tuning adjusts the parameters of a pre-trained model using a task-specific dataset to better adapt it to the target task. In our work, we fine-tuned a pre-trained neural network on a small dataset to improve classification for pulsar candidates from the MWA. We used a pre-trained network for feature extraction, taking advantage of its ability to capture hierarchical data representations. Due to the diverse data distributions of pulsar candidates from different telescopes, we provided an Application Programming Interface (API) to incorporate additional data into the fine-tuning training. By freezing the feature extraction module and fine-tuning only the final linear classifier \citep{yosinski2014transferable}, we aimed to improve its performance in identifying pulsar candidates from telescopes outside the training data.

\section{DATA} 
\label{sec:section3}
\subsection{Observational data for pipeline benchmarking}
\label{sec:section3.1}
We have compiled comprehensive datasets of pulsar candidates collected from telescopes operating at different frequencies and under varying radio environments. We train and validate the model with an expert-reviewed dataset of real pulsar samples from three telescopes (FAST, Parkes, and Arecibo). We assess the generalizability of the trained model using candidate datasets collected from the MWA and the Green Bank Telescope (GBT). Additionally, we demonstrate the speed-up of our search strategy using globular cluster data from FAST, data from a single observation of the SMART survey, and simulated data containing multiple pulsars with specified ranges of DM values.

\begin{table}
 \caption{Candidates dataset for training and validation}
 \label{tab:dataset}
 \begin{tabular}{lcc}
  \hline
  Telescope & Samples\\
  &Positive : Negative (1:1)\\
  \hline
  FAST & 3000\\
  Parkes & 5000\\
  Arecibo & 4000\\
  \hline
 \end{tabular}
\end{table}

The data from FAST, Parkes, and Arecibo were used as training and validation datasets (Table\,\ref{tab:dataset}), while observations from the MWA and GBT served as novel data environments to evaluate model generalizability. To determine whether a folded candidate is a pulsar, we perform feature identification on the two key characteristics in the candidate diagnostic plot. This supervised deep learning task requires manually annotated data. Experts inspect each candidate and label it as 1 for a true pulsar and 0 for a non-pulsar. We note that our model's reliance on time-domain features makes it susceptible to performance degradation from pulsar-like negative candidates (e.g., narrow-band RFIs). During dataset construction, we implemented manual cleansing of the training set to eliminate such RFIs.

We assembled a dataset of pulsar candidates, including both positive and negative samples, generated from five pulsar surveys processed using PRESTO. The FAST candidates were sourced from the Commensal Radio Astronomy FAST Survey (CRAFTS) \citep{li2018fast}, the Parkes candidates were collected from the High Time Resolution Universe South Low Latitude (HTRU-S LowLat) pulsar survey \citep{keith2010high}, and the Arecibo candidates from the Pulsar Arecibo L-band Feed Array (PALFA) survey \citep{cordes2006arecibo}. In addition, we collected Green Bank North Celestial Cap (GBNCC) survey candidates from the GBT \citep{stovall2014green}, and incorporated pulsar candidates from the SMART survey, conducted with the MWA.

We validated the performance of our method using real globular cluster data. Pulsar searches typically require scanning a wide range of celestial coordinates to discover new pulsars. However, within globular clusters, the detection of multiple pulsars with similar DM values becomes possible. This capability arises due to the dense stellar environment inherent to globular clusters, which provides favorable conditions for the formation and retention of pulsars, thereby facilitating the detection of multiple pulsars exhibiting similar DM values \citep{ransom2007pulsars}. We validate the feasibility of our speed-up strategy by evaluating whether all known globular cluster pulsars' ``snapshot'' candidates are retained as top-ranked candidates within narrow DM ranges.

We also evaluated the speed-up ratio using data of all detectable known pulsars from an 80-minute single observation of the SMART survey. Our test makes use of beamformed data on 43 known pulsars from observation ID 1267459328, collected during Phase II MWA’s compact configuration \citep{wayth2018phase}. We performed a large-scale blind search on these known pulsar data using PRESTO. After RFI removal, we applied de-dispersion processing over a DM range of 0-500 $\text{pc cm}^{-3}$ with a total of 6,624 trials. Then, for each trial, we apply the FFT and remove red noise. Using the parameters \text{"-zmax 100, -numharm 16 and -sigma 2.0"}, we conduct an acceleration search followed by filtering with ACCEL\_sift.py to obtain a series of potential pulsar candidates. The number of candidates generated from pointings that include known pulsars within a single observation is typically higher than that from pointings without pulsars. Therefore, if the observation field is covered with a sufficient number of pointings, the average number of candidates per pointing is unlikely to exceed that of the pointings with known pulsars. This experiment thus provides a conservative estimate of the potential speed-up achievable by our method for large-scale pulsar survey data processing. This processing pipeline allows us to evaluate the speed-up of our method within a parameter space close to that of real pulsar surveys. We applied our model to classify the ``snapshot'' candidates and selected a small subset for folding using the full data. The number of candidates in this subset serves as an indicator of the speed-up of our method.

\begin{figure} 	 
\centering
\includegraphics[width=8cm]{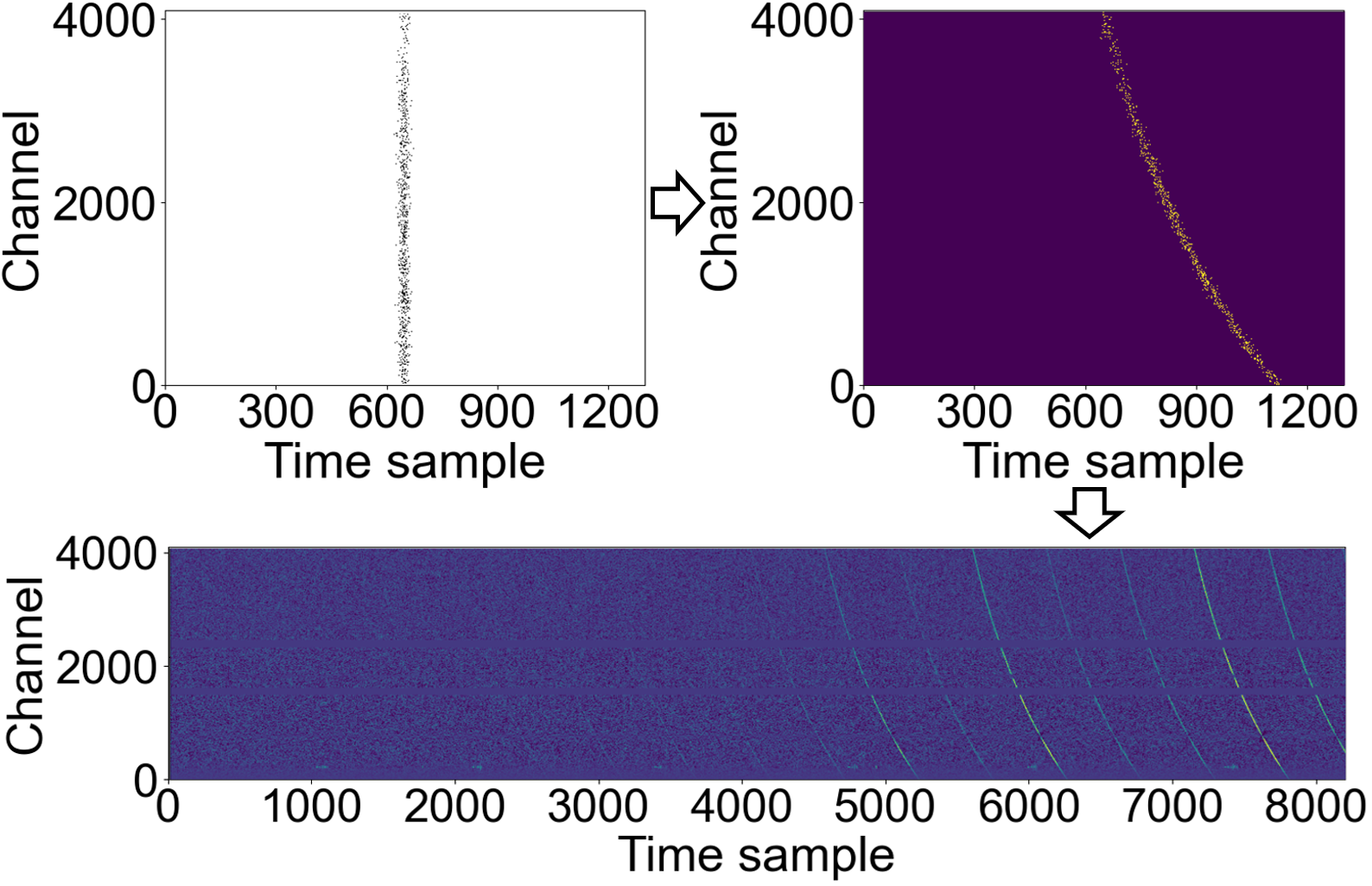}
\caption{Simulated data on pulsars were generated using our developed script that injects signals with dispersion delays into observational data. The process of simulating pulsars on FAST observational data is shown in the figure. The x-axis represents time samples, and the y-axis represents frequency channels. The simulation of pulsars involves generating single pulses with different intensities, applying time delays corresponding to the specified DM, and adding these pulses to the appropriate time samples.}
\label{fig:simulation}  
\end{figure}

\subsection{Simulated data for model performance optimization}
\label{sec:section3.2}
We developed a pulsar signal simulation script that periodically injects single pulses into PSRFITS-format files. The simulated pulse parameters are extracted from real observed single pulses, and we simulate the dispersion effect of pulsars by adding time delays to the signals. These simulated pulsars support analysis of the speed-up of our pipeline and help fine-tune our pre-trained models to better adapt to the data distribution of newly constructed telescopes.

Currently, there are several software packages that can simulate pulsar data. One such software is SIGPROC \citep{lorimer2011sigproc}, which is capable of generating filterbank format files for simulated pulsars with specified DM values and periods. One drawback of SIGPROC is its inability to generate data in the PSRFITS format. Furthermore, the method of generating Gaussian noise is inadequate for accurately simulating complex observational environments. \citet{luo2022simulating} developed an inventive software package for simulating channelized, high-time resolution radio telescope data streams. This versatile tool enables researchers to generate synthetic data tailored to specific telescope parameters and observing systems, while also supporting the injection of simulated signals into real datasets. This software lacks GPU support and cannot allow the injection of multiple pulsars in a single processing run, which makes it not well-suited for large-scale pulsar simulation in our work.

Considering the aforementioned factors, we have developed a script using PyCUDA \footnote{\url{ https://github.com/inducer/pycuda}}, specifically designed to use NVIDIA GPUs for speed-up. This script enables efficient injection of simulated pulses into observational data. As shown in Figure \ref{fig:simulation}, we extract multiple sets of parameters for the 2-D Gaussian function from real single pulses, then randomly sample one parameter set for each injected pulse and optimize it to achieve the desired SNR. Then add the specified dispersion delay to the simulated pulses and inject them into the observational data. Following this, we use the PyCUDA framework for parallel computation of dispersion delay and for parallel periodic injection of delayed pulses into observational data. 
As a result, we obtain simulated pulsar data with realistic background noise. 

Hence, it is feasible to simulate data containing a specified number of pulsars by injecting signals with distinct parameters within a defined DM range. This approach allows multiple simulated pulsars to coexist within the same DM range. By searching for pulsars within this DM range in the injected data and examining the ranking of these simulated pulsars among all candidates, we can evaluate the speed-up of our model in the folding stage of the pulsar search pipeline. Due to differences in radio environments and operating frequencies, pulsar candidates generated by different telescopes exhibit variations in data distribution. For newly constructed telescopes, they do not have enough samples to train models, so we can fine-tune the models with a small number of both real and simulated pulsar candidates, preserving the feature extraction ability obtained from training a large dataset. We validated the feasibility of the approach through fine-tuning experiments on MWA pulsar candidates.

\section{Results}
\label{sec:section4}
\subsection{Pipeline speed-up}
We benchmarked the speed-up of our strategy using 30 minutes of FAST observation data from project PT2020\_0074, targeting the globular cluster NGC 5904. The observation parameters had a sample time of $49.152\,\mathrm{\upmu s}$, 4096 frequency channels, and 4 polarizations. The 30-minute observation generates 545 GB of full data. We need to fold with all possible combinations of DM values and periods. After de-dispersion with the set of planned DM values, the size of each single de-dispersed time series is reduced by a factor of 4096 $\times$ 4, resulting in a final size of 34 MB. Following the speed-up strategy described in Section \ref{sec:section2.1}, we fold the de-dispersed time series data, with each series corresponding to a particular DM value. The time required to fold the de-dispersed time series data ($T_1$) is much shorter than the time required to fold the full data ($T_0$). The speed-up in the folding step time is given by the ratio $N_1 / N_0$,
\begin{equation}
\text{Speed-up} =
\frac{N_0 \times T_1 + N_1 \times T_0}{N_0 \times T_0}
\approx \frac{N_1}{N_0}, \quad \text{as } T_1 \ll T_0
\end{equation}
where $N_0$ is the number of DM-period combinations generated by PRESTO, and $N_1$ is the number of candidates that require folding after AI model filtering. This optimization significantly reduces the computational cost of the pulsar search by limiting full data folding, which focuses on a subset of the most promising pulsar candidates. 

In the globular cluster NGC 5904, there are seven known pulsars \citep{zhang2023discovery}. Based on their DM range from 29.31 to 30.05 $\text{pc cm}^{-3}$, the full data were de-dispersed from 29.05 to 30.25 $\text{pc cm}^{-3}$ in steps of 0.05 $\text{pc cm}^{-3}$ after careful RFI removal. We then applied acceleration search using the parameters \text{"-zmax 500, -wmax 500 and -numharm 32"}, and filtered the results with \text{ACCEL\_sift.py}, producing 1906 candidates. A total of 1264 were excluded due to DM problems (including cases with insufficient numbers across DM trials), with another 106 filtered out as harmonics, leaving 536 pulsar candidates. If we use the standard threshold of 0.5, 74 candidates were selected for further folding on the full data, including the fundamental-period and harmonic candidates of the seven known pulsars. Each known pulsar detected in the observation may correspond to multiple candidates, including both fundamental-period and harmonic ones. Our acceleration strategy performs optimally when it maximizes the filtering out of non-pulsar candidates without missing any known pulsars. To achieve this, we need the fundamental-period candidates corresponding to the known pulsars in this globular cluster to be retained after filtering by our model. Among these, the candidate with the lowest predicted probability of being a pulsar still ranks within the top 13\% of all candidates, corresponding to the top 72 candidates. This shows that the fundamental-period and harmonic candidates of known pulsars are ranked near the top of all candidates, ensuring that our strategy reduces the number of folding parameters while avoiding any missed pulsars. After using our AI model, the number of folding parameters was reduced by a factor of 10, resulting in a tenfold speed-up of the folding process. Thus, AI effectively speeds up the folding process tenfold while successfully identifying all detectable pulsars in the global cluster. 

We used data from 43 known pulsars in a single observation of the SMART survey to evaluate the speed-up across a large parameter space. Using the search pipeline described in Section \ref{sec:section3.1}, we processed the data and obtained 25,691 potential DM-period combinations. These combinations were then folded on 1-D time series data to generate the "snapshot"  candidates. We applied a standard threshold of 0.5 to identify and select candidates. Our model selected 436 candidates for further folding with the full data. Manual inspection of these selected candidates confirmed that they include the fundamental candidates of all 43 known pulsars, as well as some harmonic candidates. As mentioned in Section \ref{sec:section3.2}, pointings containing known pulsars within a single observation typically produce more candidates than those without pulsars. Therefore, the average number of candidates from searches on known pulsar data tends to be somewhat higher than the average number generated across all pointings in the survey. Using the same search pipeline, we performed a blind search on 960 pointings from the same observation and obtained a total of 569,869 candidates. Each pointing containing a known pulsar produced an average of 597 candidates, slightly higher than the 593 candidates from other pointings. Among the candidates generated from these 960 pointings, our AI model selected 6,478, with only 1.14\% marked for further verification. So, our speed-up tests using known pulsar data from a single observation in the SMART survey provide a relatively conservative estimate. Consistent with the previous analysis on globular cluster data, this result suggests that our method can accelerate the folding step by approximately a factor of 60 in large-scale pulsar searches conducted with the MWA-SMART survey. This experiment shows significantly better performance compared to the globular cluster test. This improvement is mainly due to the excellent radio environment of MWA \citep{offringa2015low}, which results in much less RFI than FAST. In addition, the globular cluster test was conducted in a restricted parameter space that contained multiple known pulsars, leading to a higher number of pulsar-like candidates. Considering the above analysis, our method performs better in the large-scale parameter space typical of pulsar surveys.

We applied the simulation program described in Section \ref{sec:section3.2} to 10 minutes of data from the FAST pulsar backend, while the telescope was not pointing to any known pulsars. This data served as the background for periodically injecting pulses to simulate pulsars. We generate 20 simulated pulsars with the following injected parameters: the DM ranges from 91 to 109 $\text{pc cm}^{-3}$, the SNR ranges from 5 to 15, and the duty cycle ranges from 1\% to 10\%. We evaluated the model on simulated data to verify the tenfold speed-up during the folding process. Using PRESTO with 0.1 DM value increments between 90 and 110 $\text{pc cm}^{-3}$ generated 200 de-dispersion trails. The application of the acceleration search parameters \text{"-zmax 200, -wmax 200 and -numharm 16"}, followed by filtering with \text{ACCEL\_sift.py}, found 2141 candidates. After removing 1697 candidates due to DM problems and excluding 143 harmonics, 299 pulsar candidates remained. Among them, 48 candidates had predicted pulsar probabilities above 50\% from the classification model, successfully recovering all 20 simulated pulsars. And the 20 injected pulsars ranked within the top 22 scores, representing the top 7.4\% of all pulsar candidates. Using FAST observational globular cluster data and simulated data, our trained model consistently ranked known pulsars within the top \text{10\%} of all candidates. This demonstrated that our strategy recovered all known pulsars and achieved a tenfold speed-up in the folding step within a restricted parameter space.


\begin{table}
  \centering
  \caption{Confusion matrix for pulsar classification.}
  \label{tab:confusion}
  \setlength{\tabcolsep}{6pt}
  \renewcommand{\arraystretch}{1.2}
  \begin{tabular}{lcc}
    \hline
    \textbf{} & \textbf{Predicted: Pulsar} & \textbf{Predicted: Non-pulsar} \\
    \hline
    \textbf{Actual: Pulsar}     & True Positive (TP)  & False Negative (FN) \\
    \textbf{Actual: Non-pulsar} & False Positive (FP) & True Negative (TN) \\
    \hline
  \end{tabular}
\end{table}

\subsection{Evaluation of AI model performance}
We used the following four metrics to evaluate the model's performance for pulsar classification, calculating them based on the confusion matrix presented in Table~\ref{tab:confusion}.
\begin{enumerate}
 \item Accuracy measures the overall correctness of the model by calculating the proportion of true results (both true positives and true negatives) among the total number of cases. It is defined as the ratio of correctly predicted instances to the total instances. 
 
 Accuracy = $\frac{\text{TP} + \text{TN}}{\text{TP} + \text{TN} + \text{FP} + \text{FN}}$
 \item Precision indicates the proportion of positive identifications that were actually correct. It is defined as the ratio of true positive results to the sum of true positives and false positives.
 
 Precision = $\frac{\text{TP}}{\text{TP} + \text{FP}}$
 \item Recall measures the proportion of actual positives that were correctly identified by the model. It is defined as the ratio of true positive results to the sum of true positives and false negatives.
 
 Recall = $\frac{\text{TP}}{\text{TP} + \text{FN}}$
 
 \item F-score (F1) is the harmonic mean of precision and recall. It provides a metric that balances both the precision and recall of a model.

 F-score = $2 \times \frac{\text{Precision} \times \text{Recall}}{\text{Precision} + \text{Recall}}$
\end{enumerate}

The evaluation protocol comprised three distinct phases: Phase I assessed the model's performance on 3,000 candidates from an independent validation subset (25\%) of the complete datasets (12,000 samples) from FAST/Arecibo/Parkes. Phase II conducted blind testing with 10,000 fundamental candidates (4,000 positive and 6,000 negative) from the MWA-SMART pulsar survey, where the 4,000 positive candidates are essentially multiple detections of 86 known pulsars, as a consequence of the MWA complex tied-array beam patterns with multiple strong side lobes \citep{bhat2023southern}. The results demonstrate maintained detection efficacy (96.65\% recall) in independent test datasets. After fine-tuning with 1,000 samples from MWA, the recall increased to 97.925\%. Phase III evaluated the GBNCC survey candidates (from \citeauthor{zhu2014searching} \citeyear{zhu2014searching}) using our trained model with cross-validation, comparing the results with prior work. All phases applied consistent pre-processing pipelines and selection criteria, ensuring comparability across different survey configurations. 

In Phase II and Phase III, we evaluate classification performance using our model (\textsc{Multi}) and the \textsc{PICS} model \citep{zhu2014searching}, with both models applying thresholds of 0.5 and 0.3. A threshold of 0.5 is used in binary classification to separate the two classes based on equal probability. Since non-pulsar candidates greatly exceed pulsar candidates in pulsar surveys, and our method serves as an initial filtering step where missing pulsar candidates are undesirable, we also test a lower threshold of 0.3. The PICS model is tested with the recommended \text{"clfl2\_PALFA.pkl"} weights. Since PICS adopts an ensemble structure that independently trains on four features (profile, time-phase, frequency-phase, and DM curve) and combines their outputs using logistic regression \citep{hosmer2013applied}, we design a comparison experiment as follows. In Phase II, we average the outputs of PICS's ANN (for profile) and CNN (for time-phase) to represent its prediction based on time-domain features, and compare the results with \textsc{Multi} on SMART survey candidates. In Phase III, PICS uses all four features for classification, while \textsc{Multi} relies on time-domain features, allowing us to analyze the impact of using temporal information alone.


\begin{table}
\caption{Performance metrics for different telescopes (accuracy, precision, recall, F1).}
\setlength{\tabcolsep}{5.5pt}
\label{tab:training_evaluation}
\begin{tabular}{|c|c|c|c|c|}
\hline
\textbf{Telescope} & \textbf{Accuracy (\%)} & \textbf{Precision (\%)} & \textbf{Recall (\%)} & \textbf{F1 (\%)} \\
\hline
FAST & 98.17 & 99.22 & 96.60 & 97.90 \\
Parkes & 98.11 & 98.03 & 98.60 & 98.31 \\
Arecibo & 98.73 & 98.10 & 99.28 & 98.69 \\
\hline
\textbf{Overall} & \textbf{98.30} & \textbf{98.25} & \textbf{98.44} & \textbf{98.34} \\
\hline
\end{tabular}
\end{table}

\subsubsection{Evaluation of validation data}

As shown in Table~\ref{tab:training_evaluation}, the internal validation achieved 98.3\% accuracy on the 3,000 candidates test subset, with precision and recall rates of 98.25\% and 98.44\%, respectively. Notably, the model demonstrated robust performance across different telescopes within the training data: FAST (96.6\% recall), Parkes (98.6\% recall), and Arecibo (99.28\% recall). This consistency suggests effective generalization across varying observational parameters and instrumental characteristics.

\subsubsection{Model evaluation with MWA data}

Table~\ref{tab:smart} summarizes the results of the SMART survey. Both models use time-domain features for classification. The numbers indicate the counts of correctly classified positive samples and misclassified negative samples. The PICS model downsampling algorithm standardizes the feature dimensions by reducing the profile features to 64 and the time-phase features to 48. This leads to a decrease in precision for candidates from the SMART survey, which originally had a bin number of 100. Consequently, our pulsar time-domain feature classification model, designed for multi-dimensional input, effectively handles the varying number of bins among the original candidates. Due to the excellent radio environment of MWA \citep{offringa2015low}, the number of pulsar-like negative candidates caused by narrow-band RFIs is very low. \textsc{Multi} model exhibits consistently strong performance on both positive and negative pulsar candidates. 

For the SMART survey, the classification performance is further improved through the fine-tuning strategy described in Section \ref{sec:section2.3}. A fine-tuning dataset is constructed using 200 pulsars and 500 non-pulsars from the SMART survey, added with 300 simulated pulsars generated based on MWA data, resulting in a total of 1,000 samples. Fine-tuning is performed on the final classification layer to adapt the model to the data distribution of SMART pulsar candidates, enhancing its discrimination capability for both positive and negative samples. The remaining misclassified pulsars are mostly candidates generated from bright sources observed through MWA sidelobes. These candidates are characterized by faint, diffuse vertical structures in their time-phase features, appearing at a fixed pulse phase and confined to a limited portion of the time axis. These structures correspond to periodic pulses occurring during part of the observation, caused by bright pulsars observed via the sidelobes of the beamformed pointing. In the SMART test dataset, all known pulsars were correctly identified by the \textsc{Multi} model. Only a small number of positive samples observed through sidelobe responses were not identified.

\subsubsection{GBT data comparison}

The results from the GBT are summarized in the following Table~\ref{tab:gbncc}. The PICS model considers both time and frequency features, while \textsc{Multi} focuses only on time-domain features. For the GBNCC dataset, the \textsc{Multi} model effectively identifies the intrinsic features of both fundamental and harmonic candidates. However, the model's capacity to filter out RFI is significantly impeded by the lack of detailed frequency and DM information. In the future, we will train a model using advanced algorithms to identify candidates with detailed time and frequency information, enhancing RFI filtering.



\begin{table}
  \centering
  \caption{SMART results}
  \label{tab:smart}
  \setlength{\tabcolsep}{5.4pt}
  \renewcommand{\arraystretch}{1.2}
  \begin{tabular}{lccc}
    \hline
    \textbf{Model} & \textbf{Threshold} & \textbf{Pulsar (4,000)} & \textbf{Non-pulsar (6,000)} \\
    \hline
    \textsc{PICS}             & >0.5 & 3587 & 362 \\
                     & >0.3 & 3789 & 931 \\
    \hline
    \textsc{Multi}            & >0.5 & 3866 & 270 \\
                     & >0.3 & 3910 & 360 \\
    \hline
    Finetuned \textsc{Multi}  & >0.5 & 3917 & 203 \\
                     & >0.3 & 3940 & 300 \\
    \hline
  \end{tabular}
\end{table}


\begin{table}
  \centering
  \caption{GBNCC results}
  \label{tab:gbncc}
  \setlength{\tabcolsep}{2.3pt}
  \renewcommand{\arraystretch}{1.2}
  \begin{tabular}{lccc}
    \hline
    \textbf{Model} & \textbf{Fundamental (56)} & \textbf{Harmonic (221)} & \textbf{Non-pulsar (10,000)} \\
    \hline
    \textsc{PICS} (>0.5)   & 53 & 207 & 43 \\
    \textsc{PICS} (>0.3)   & 53 & 211 & 132 \\
    \textsc{Multi} (>0.5)  & 56 & 219 & 762 \\
    \textsc{Multi} (>0.3)  & 56 & 221 & 1077 \\
    \hline
  \end{tabular}
\end{table}

\section{Discussion}
\label{sec:section5}
The SKA, as the next-generation radio telescope, is predicted to detect over 20,000 new pulsars, of which 7,000-9,000 of them from SKA-1 Low \citep{keane2014cosmic,xue2017census}, fundamentally advancing our understanding of neutron star populations and extreme astrophysical phenomena. Due to the large field of view (FoV) of radio telescope arrays, even with long single observation times used in pulsar surveys, it is possible to complete the pulsar survey plan in a short amount of time compared to single-dish telescopes. For example, each observation in the SMART survey uses an 80-minute dwell time, allowing the full survey coverage to be completed in less than 100 hours of telescope time \citep{bhat2023southern}. As the observation time for a single point is extended, the resulting full data size increases correspondingly. This directly increases the time required to fold the increasing size of the full data. Note that our method effectively accelerates pulsar searches if folding dominates computational time. Our work addresses this critical bottleneck in pulsar search pipelines through three key innovations.

First, we develop a time-domain feature extraction method from pre-folded "snapshot" candidates that enables pulsar identification with 98.44\% recall before full folding. This performance is validated on both cross-validation and independent holdout datasets. This early-stage filtering reduces the required full-fold operations by 90\% in real and simulated data within a restricted parameter space. On a large parameter space approximating the processing in the SMART survey, our method improved the efficiency of the folding step by a factor of 60. This corresponds to a 98\% reduction in required full-fold operations.

Second, we developed an efficient pulsar simulation script that injects simulated pulsars with diverse parameters into real observational noise data, addressing the data scarcity challenge for newly built telescopes. This approach effectively resolves data distribution discrepancies across candidates from different telescopes, particularly when insufficient real pulsar samples are available for model fine-tuning. 

Third, the hybrid network architecture demonstrates robust adaptability to varying input dimensions. Our denoising network further enhances performance by reinforcing 2-D features. When processing pulsar candidates with different bin sizes, the model maintains >98\% classification accuracy through the learnable feature layers. 

The MWA-SMART survey demonstrates the advantage of a large FoV combined with extended single observation times, achieving a 20-fold increase in dwell time away from the Galactic plane compared to the HTRU survey \citep{bhat2023southern,keith2010high}. The future SKA survey will also benefit from a relatively large FoV and longer observations, making it critical to reduce computational overhead in the folding step. By fine-tuning the model using the simulated pulsar samples from the newly constructed radio telescope, a robust model can be established during the preliminary investigation phase of the pulsar survey. The methodology establishes a reliable pathway for deploying an AI model in pulsar surveys of next-generation telescopes such as the SKA. 


\section*{Acknowledgements}

We sincerely thank the referee, Scott Ransom, for his highly insightful and constructive comments which were so valuable in helping us improve the manuscript. This work is supported by the National SKA Program of China No. 2020SKA0120200, the National Natural Science Foundation of China (NSFC) grant Nos. 12421003, 12503055, 12041303, 62176247, 12203070, 12041304, the Beijing Nova Program (No. 20250484786), the Postdoctoral Fellowship Program of CPSF under Grant Number GZB20250737. This scientific work uses data obtained from Inyarrimanha Ilgari Bundara, the CSIRO Murchison Radio-astronomy Observatory. We acknowledge the Wajarri Yamaji People as the Traditional Owners and Native Title Holders of the observatory site. 
Support for the operation of the MWA is provided by the Australian Government (NCRIS), under a contract to Curtin University administered by Astronomy Australia Limited (AAL). 
We acknowledge the Pawsey Supercomputing Centre which is supported by the Western Australian and Australian Governments. This work made use of the data from FAST (Five-hundred-meter Aperture Spherical radio Telescope).  FAST is a Chinese national mega-science facility, operated by National Astronomical Observatories, Chinese Academy of Sciences.

\section*{Code and Data Availability}

The data supporting this article will be shared when a reasonable request is made to the corresponding author. The code used for model implementation and experiments in this study is publicly available at: https://github.com/ifuqy/Multi. The repository also provides the trained model weights used in our experiments, along with usage instructions.



\bibliographystyle{mnras}
\bibliography{example} 

@article{hulse1975discovery,
  title={Discovery of a pulsar in a binary system},
  author={Hulse, Russell A and Taylor, Joseph H},
  journal={Astrophysical Journal, vol. 195, Jan. 15, 1975, pt. 2, p. L51-L53.},
  volume={195},
  pages={L51--L53},
  year={1975}
}

@article{manchester2005australia,
  title={The Australia telescope national facility pulsar catalogue},
  author={Manchester, R Nꎬ and Hobbs, G Bꎬ and Teoh, Aꎬ and Hobbs, M\_},
  journal={The Astronomical Journal},
  volume={129},
  number={4},
  pages={1993},
  year={2005},
  publisher={IOP Publishing}
}

@article{manchester1978second,
  title={The second Molonglo pulsar survey--discovery of 155 pulsars},
  author={Manchester, RN and Lyne, AG and Taylor, JH and Durdin, JM and Large, MI and Little, AG},
  journal={Monthly Notices of the Royal Astronomical Society},
  volume={185},
  number={2},
  pages={409--421},
  year={1978},
  publisher={Oxford University Press Oxford, UK}
}

@article{zhu2014searching,
  title={Searching for pulsars using image pattern recognition},
  author={Zhu, W., W. and Berndsen, Aaron and Madsen, EC and Tan, M and Stairs, IH and Brazier, A and Lazarus, P and Lynch, R and Scholz, P and Stovall, K and others},
  journal={The Astrophysical Journal},
  volume={781},
  number={2},
  pages={117},
  year={2014},
  publisher={IOP Publishing}
}

@article{balakrishnan2021pulsar,
  title={Pulsar candidate identification using semi-supervised generative adversarial networks},
  author={Balakrishnan, Vishnu and Champion, David and Barr, Ewan and Kramer, Michael and Sengar, Rahul and Bailes, Matthew},
  journal={Monthly Notices of the Royal Astronomical Society},
  volume={505},
  number={1},
  pages={1180--1194},
  year={2021},
  publisher={Oxford University Press}
}

@article{cooley1965algorithm,
  title={An algorithm for the machine calculation of complex Fourier series},
  author={Cooley, James W and Tukey, John W},
  journal={Mathematics of computation},
  volume={19},
  number={90},
  pages={297--301},
  year={1965}
}

@article{rickett1990radio,
  title={Radio propagation through the turbulent interstellar plasma},
  author={Rickett, Barney J},
  journal={Annual review of astronomy and astrophysics},
  volume={28},
  number={1},
  pages={561--605},
  year={1990},
  publisher={Annual Reviews 4139 El Camino Way, PO Box 10139, Palo Alto, CA 94303-0139, USA}
}

@book{lorimer2005handbook,
  title={Handbook of pulsar astronomy},
  author={Lorimer, Duncan Ross and Kramer, Michael},
  volume={4},
  year={2005},
  series = {Cambridge Observing Handbooks for Research Astronomers},
  publisher={Cambridge university press}
}

@article{taylor1991millisecond,
  title={Millisecond pulsars: Nature's most stable clocks},
  author={Taylor, Joseph H},
  journal={Proceedings of the IEEE},
  volume={79},
  number={7},
  pages={1054--1062},
  year={1991},
  publisher={IEEE}
}

@article{luo2022simulating,
  title={Simulating high-time resolution radio-telescope observations},
  author={Luo, Rui and Hobbs, George and Yong, Suk Yee and Zic, Andrew and Toomey, Lawrence and Dai, Shi and Dunning, Alex and Li, Di and Marshman, Tommy and Wang, Chen and others},
  journal={Monthly Notices of the Royal Astronomical Society},
  volume={513},
  number={4},
  pages={5881--5891},
  year={2022},
  publisher={Oxford University Press}
}

@article{cordes2006arecibo,
  title={Arecibo pulsar survey using ALFA. I. Survey strategy and first discoveries},
  author={Cordes, James M and Freire, Paulo Cesar Carvalho and Lorimer, Duncan R and Camilo, Fernando and Champion, David J and Nice, David J and Ramachandran, Ramesh and Hessels, JWT and Vlemmings, Wouter and Van Leeuwen, Joeri and others},
  journal={The Astrophysical Journal},
  volume={637},
  number={1},
  pages={446},
  year={2006},
  publisher={IOP Publishing}
}

@inproceedings{he2016deep,
  title={Deep residual learning for image recognition},
  author={He, Kaiming and Zhang, Xiangyu and Ren, Shaoqing and Sun, Jian},
  booktitle={Proceedings of the IEEE conference on computer vision and pattern recognition},
  pages={770--778},
  year={2016}
}

@article{lecun1998gradient,
  title={Gradient-based learning applied to document recognition},
  author={LeCun, Yann and Bottou, L{\'e}on and Bengio, Yoshua and Haffner, Patrick},
  journal={Proceedings of the IEEE},
  volume={86},
  number={11},
  pages={2278--2324},
  year={1998},
  publisher={Ieee}
}

@inproceedings{woo2018cbam,
  title={Cbam: Convolutional block attention module},
  author={Woo, Sanghyun and Park, Jongchan and Lee, Joon-Young and Kweon, In So},
  booktitle={Proceedings of the European conference on computer vision (ECCV)},
  pages={3--19},
  year={2018}
}

@inproceedings{ioffe2015batch,
  title={Batch normalization: Accelerating deep network training by reducing internal covariate shift},
  author={Ioffe, Sergey and Szegedy, Christian},
  booktitle={International conference on machine learning},
  pages={448--456},
  year={2015},
  organization={pmlr}
}

@inproceedings{nair2010rectified,
  title={Rectified linear units improve restricted boltzmann machines},
  author={Nair, Vinod and Hinton, Geoffrey E},
  booktitle={Proceedings of the 27th international conference on machine learning (ICML-10)},
  pages={807--814},
  year={2010}
}

@article{li2018fast,
  title={FAST in space: considerations for a multibeam, multipurpose survey using China's 500-m aperture spherical radio telescope (FAST)},
  author={Li, Di and Wang, Pei and Qian, Lei and Krco, Marko and Dunning, Alex and Jiang, Peng and Yue, Youling and Jin, Chenjin and Zhu, Yan and Pan, Zhichen and others},
  journal={IEEE Microwave Magazine},
  volume={19},
  number={3},
  pages={112--119},
  year={2018},
  publisher={IEEE}
}

@article{keith2010high,
  title={The high time resolution universe pulsar survey--i. system configuration and initial discoveries},
  author={Keith, MJ and Jameson, A and Van Straten, W and Bailes, M and Johnston, S and Kramer, M and Possenti, A and Bates, SD and Bhat, NDR and Burgay, M and others},
  journal={Monthly Notices of the Royal Astronomical Society},
  volume={409},
  number={2},
  pages={619--627},
  year={2010},
  publisher={The Royal Astronomical Society}
}

@article{stovall2014green,
  title={The green bank northern celestial cap pulsar survey. I. Survey description, data analysis, and initial results},
  author={Stovall, K and Lynch, RS and Ransom, SM and Archibald, AM and Banaszak, S and Biwer, CM and Boyles, J and Dartez, LP and Day, D and Ford, AJ and others},
  journal={The Astrophysical Journal},
  volume={791},
  number={1},
  pages={67},
  year={2014},
  publisher={IOP Publishing}
}

@article{bhat2023southern,
  title={The Southern-sky MWA Rapid Two-metre (SMART) pulsar survey—I. Survey design and processing pipeline},
  author={Bhat, NDR and Swainston, NA and McSweeney, SJ and Xue, M and Meyers, BW and Kudale, S and Dai, S and Tremblay, SE and van Straten, W and Shannon, RM and others},
  journal={Publications of the Astronomical Society of Australia},
  volume={40},
  pages={e021},
  year={2023},
  publisher={Cambridge University Press}
}

@inproceedings{ronneberger2015u,
  title={U-net: Convolutional networks for biomedical image segmentation},
  author={Ronneberger, Olaf and Fischer, Philipp and Brox, Thomas},
  booktitle={Medical image computing and computer-assisted intervention--MICCAI 2015: 18th international conference, Munich, Germany, October 5-9, 2015, proceedings, part III 18},
  pages={234--241},
  year={2015},
  organization={Springer}
}

@article{ho2020denoising,
  title={Denoising diffusion probabilistic models},
  author={Ho, Jonathan and Jain, Ajay and Abbeel, Pieter},
  journal={Advances in neural information processing systems},
  volume={33},
  pages={6840--6851},
  year={2020}
}

@article{lorimer2011sigproc,
  title={SIGPROC: pulsar signal processing programs},
  author={Lorimer, DR},
  journal={Astrophysics Source Code Library},
  pages={ascl--1107},
  year={2011}
}

@article{ransom2011presto,
  title={PRESTO: pulsar exploration and search toolkit},
  author={Ransom, Scott},
  journal={Astrophysics source code library},
  pages={ascl--1107},
  year={2011}
}

@article{hotan2004psrchive,
  title={PSRCHIVE and PSRFITS: an open approach to radio pulsar data storage and analysis},
  author={Hotan, Aidan W and van Straten, Willem and Manchester, Richard N},
  journal={Publications of the Astronomical Society of Australia},
  volume={21},
  number={3},
  pages={302--309},
  year={2004},
  publisher={Cambridge University Press}
}

@article{eatough2010selection,
  title={Selection of radio pulsar candidates using artificial neural networks},
  author={Eatough, R Pꎬ and Molkenthin, N and Kramer, M and Noutsos, A and Keith, MJ and Stappers, BW and Lyne, AG},
  journal={Monthly Notices of the Royal Astronomical Society},
  volume={407},
  number={4},
  pages={2443--2450},
  year={2010},
  publisher={Blackwell Publishing Ltd Oxford, UK}
}

@article{lee2013peace,
  title={PEACE: pulsar evaluation algorithm for candidate extraction--a software package for post-analysis processing of pulsar survey candidates},
  author={Lee, KJ and Stovall, K and Jenet, FA and Martinez, J and Dartez, LP and Mata, A and Lunsford, G and Cohen, S and Biwer, CM and Rohr, M and others},
  journal={Monthly Notices of the Royal Astronomical Society},
  volume={433},
  number={1},
  pages={688--694},
  year={2013},
  publisher={The Royal Astronomical Society}
}

@article{barr2013northern,
  title={The northern high time resolution universe pulsar survey--I. Setup and initial discoveries},
  author={Barr, Ewan D and Champion, David J and Kramer, Michael and Eatough, Ralph P and Freire, Paulo CC and Karuppusamy, Ramesh and Lee, KJ and Verbiest, Joris PW and Bassa, Cees G and Lyne, Andrew G and others},
  journal={Monthly Notices of the Royal Astronomical Society},
  volume={435},
  number={3},
  pages={2234--2245},
  year={2013},
  publisher={Oxford University Press}
}

@article{keith2009discovery,
  title={Discovery of 28 pulsars using new techniques for sorting pulsar candidates},
  author={Keith, MJ and Eatough, RP and Lyne, AG and Kramer, M and Possenti, A and Camilo, F and Manchester, RN},
  journal={Monthly Notices of the Royal Astronomical Society},
  volume={395},
  number={2},
  pages={837--846},
  year={2009},
  publisher={Blackwell Publishing Ltd Oxford, UK}
}

@article{zhang2023discovery,
  title={Discovery and Timing of Millisecond Pulsars in the Globular Cluster M5 with FAST and Arecibo},
  author={Zhang, Lei and Freire, Paulo CC and Ridolfi, Alessandro and Pan, Zhichen and Zhao, Jiaqi and Heinke, Craig O and Chen, Jianxing and Cadelano, Mario and Pallanca, Cristina and Hou, Xian and others},
  journal={The Astrophysical Journal Supplement Series},
  volume={269},
  number={2},
  pages={56},
  year={2023},
  publisher={IOP Publishing}
}

@article{cai2023pulsar,
  title={Pulsar candidate classification using a computer vision method from a combination of convolution and attention},
  author={Cai, Nannan and Han, Jinlin and Jing, Weicong and Zhang, Zekai and Zhou, Dejiang and Chen, Xue},
  journal={Research in Astronomy and Astrophysics},
  volume={23},
  number={10},
  pages={104005},
  year={2023},
  publisher={IOP Publishing}
}

@inproceedings{szegedy2015going,
  title={Going deeper with convolutions},
  author={Szegedy, Christian and Liu, Wei and Jia, Yangqing and Sermanet, Pierre and Reed, Scott and Anguelov, Dragomir and Erhan, Dumitru and Vanhoucke, Vincent and Rabinovich, Andrew},
  booktitle={Proceedings of the IEEE conference on computer vision and pattern recognition},
  pages={1--9},
  year={2015}
}

@article{offringa2015low,
  title={The low-frequency environment of the Murchison Widefield Array: radio-frequency interference analysis and mitigation},
  author={Offringa, AR and Wayth, RB and Hurley-Walker, N and Kaplan, DL and Barry, N and Beardsley, AP and Bell, ME and Bernardi, GIANNI and Bowman, JD and Briggs, F and others},
  journal={Publications of the Astronomical Society of Australia},
  volume={32},
  pages={e008},
  year={2015},
  publisher={Cambridge University Press}
}

@book{jahne2005digital,
  title={Digital image processing},
  author={J{\"a}hne, Bernd},
  year={2005},
  publisher={Springer}
}

@book{hosmer2013applied,
  title={Applied logistic regression},
  author={Hosmer Jr, David W and Lemeshow, Stanley and Sturdivant, Rodney X},
  year={2013},
  publisher={John Wiley \& Sons}
}

@article{keane2014cosmic,
  title={A cosmic census of radio pulsars with the SKA},
  author={Keane, EF and Bhattacharyya, B and Kramer, M and Stappers, BW and Bates, SD and Burgay, M and Chatterjee, S and Champion, DJ and Eatough, RP and Hessels, JWT and others},
  journal={arXiv preprint arXiv:1501.00056},
  year={2014}
}

@article{1968Natur.217..709H,
  title={Observation of a Rapidly Pulsating Radio Source},
  author={Hewish, A and Bell, SJ and Pilkington, JDH and Scott, PF and Collins, RA},
  journal={Nature},
  volume = {217},
  number = {5130},
  pages = {709-713},
  year={1968}
}

@article{xue2017census,
  title={A Census of Southern Pulsars at 185 MHz},
  author={Xue, Mengyao and Bhat, NDR and Tremblay, SE and Ord, SM and Sobey, C and Swainston, NA and Kaplan, DL and Johnston, Simon and Meyers, BW and McSweeney, SJ},
  journal={Publications of the Astronomical Society of Australia},
  volume={34},
  pages={e070},
  year={2017},
  publisher={Cambridge University Press}
}

@article{tingay2013murchison,
  title={The Murchison widefield array: the square kilometre array precursor at low radio frequencies},
  author={Tingay, Steven John and Goeke, Robert and Bowman, Judd D and Emrich, David and Ord, Stephen M and Mitchell, Daniel A and Morales, Miguel F and Booler, Tom and Crosse, Brian and Wayth, Randall B and others},
  journal={Publications of the Astronomical Society of Australia},
  volume={30},
  pages={e007},
  year={2013},
  publisher={Cambridge University Press}
}

@article{wayth2018phase,
  title={The phase II Murchison widefield array: design overview},
  author={Wayth, Randall B and Tingay, Steven J and Trott, Cathryn M and Emrich, David and Johnston-Hollitt, Melanie and McKinley, Ben and Gaensler, BM and Beardsley, AP and Booler, T and Crosse, B and others},
  journal={Publications of the Astronomical Society of Australia},
  volume={35},
  pages={e033},
  year={2018},
  publisher={Cambridge University Press}
}

@article{tan2018ensemble,
  title={Ensemble candidate classification for the LOTAAS pulsar survey},
  author={Tan, Chia Min and Lyon, RJ and Stappers, BW and Cooper, Sally and Hessels, JWT and Kondratiev, VI and Michilli, Daniele and Sanidas, Sotiris},
  journal={Monthly Notices of the Royal Astronomical Society},
  volume={474},
  number={4},
  pages={4571--4583},
  year={2018},
  publisher={Oxford University Press}
}

@article{sanidas2019lofar,
  title={The LOFAR tied-array all-sky survey (LOTAAS): survey overview and initial pulsar discoveries},
  author={Sanidas, S and Cooper, S and Bassa, CG and Hessels, JWT and Kondratiev, VI and Michilli, D and Stappers, BW and Tan, CM and Van Leeuwen, J and Cerrigone, L and others},
  journal={Astronomy \& Astrophysics},
  volume={626},
  pages={A104},
  year={2019},
  publisher={EDP Sciences}
}

@article{han2021fast,
  title={The FAST Galactic Plane Pulsar Snapshot survey: I. Project design and pulsar discoveries⋆},
  author={Han, JL and Wang, Chen and Wang, PF and Wang, Tao and Zhou, DJ and Sun, Jing-Hai and Yan, Yi and Su, Wei-Qi and Jing, Wei-Cong and Chen, Xue and others},
  journal={Research in Astronomy and Astrophysics},
  volume={21},
  number={5},
  pages={107},
  year={2021},
  publisher={IOP Publishing}
}

@article{jiang2020fundamental,
  title={The fundamental performance of FAST with 19-beam receiver at L band},
  author={Jiang, Peng and Tang, Ning-Yu and Hou, Li-Gang and Liu, Meng-Ting and Kr{\v{c}}o, Marko and Qian, Lei and Sun, Jing-Hai and Ching, Tao-Chung and Liu, Bin and Duan, Yan and others},
  journal={Research in Astronomy and Astrophysics},
  volume={20},
  number={5},
  pages={064},
  year={2020},
  publisher={IOP Publishing}
}

@article{yosinski2014transferable,
  title={How transferable are features in deep neural networks?},
  author={Yosinski, Jason and Clune, Jeff and Bengio, Yoshua and Lipson, Hod},
  journal={Advances in neural information processing systems},
  volume={27},
  year={2014}
}

@article{ransom2007pulsars,
  title={Pulsars in globular clusters},
  author={Ransom, Scott M},
  journal={Proceedings of the International Astronomical Union},
  volume={3},
  number={S246},
  pages={291--300},
  year={2007},
  publisher={Cambridge University Press}
}

@inproceedings{hu2018squeeze,
  title={Squeeze-and-excitation networks},
  author={Hu, Jie and Shen, Li and Sun, Gang},
  booktitle={Proceedings of the IEEE conference on computer vision and pattern recognition},
  pages={7132--7141},
  year={2018}
}

@inproceedings{vincent2008extracting,
  title={Extracting and composing robust features with denoising autoencoders},
  author={Vincent, Pascal and Larochelle, Hugo and Bengio, Yoshua and Manzagol, Pierre-Antoine},
  booktitle={Proceedings of the 25th international conference on Machine learning},
  pages={1096--1103},
  year={2008}
}

@article{krizhevsky2012imagenet,
  title={Imagenet classification with deep convolutional neural networks},
  author={Krizhevsky, Alex and Sutskever, Ilya and Hinton, Geoffrey E},
  journal={Advances in neural information processing systems},
  volume={25},
  year={2012}
}

@article{jing2025fast,
  title={FAST Pulsar Database: II. Scattering profiles of 122 Pulsars},
  author={Jing, WC and Han, JL and Wang, C and Wang, PF and Wang, T and Cai, NN and Xu, J and Yang, ZL and Zhou, DJ and Yan, Yi and others},
  journal={arXiv preprint arXiv:2506.14519},
  year={2025}
}




\appendix
\section{model details}
\label{sec:advanced}

The Peak Signal-to-Noise Ratio (PSNR) is computed as:
\noindent 
\begin{equation}
\begin{aligned}
\text{PSNR} &= 20 \cdot \log_{10} \left( 
\frac{
\text{MAX}
}{
\sqrt{\text{MSE} + \varepsilon}
}
\right) \\
\text{MAX} &= \max(\text{pred}_{\max}, \text{orig}_{\max}) - \min(\text{pred}_{\min}, \text{orig}_{\min}) \\
\text{MSE} &= \frac{1}{N} \sum_{i=1}^{N} (\text{pred}_i - \text{orig}_i)^2
\end{aligned}
\end{equation}

where $\text{MAX}$ is the dynamic range of the images (i.e., the difference between the maximum and minimum pixel values), $\text{MSE}$ is the mean squared error between the predicted and target images, and $\varepsilon$ is a small constant 
(e.g., $\varepsilon = 10^{-8}$) added for numerical stability to avoid division by zero.

\vspace{1em}
\noindent The loss function of the denoising model is defined as:
\noindent 
\begin{equation}
\begin{aligned}
\mathcal{L}_{\text{total}} &= \lambda_{\text{MSE}} \cdot \mathcal{L}_{\text{MSE}} + 
\lambda_{\text{SSIM}} \cdot \mathcal{L}_{\text{SSIM}} + 
\lambda_{\text{grad}} \cdot \mathcal{L}_{\text{grad}} \\
\lambda_{\text{MSE}} &= 0.6, \quad
\lambda_{\text{SSIM}} = 0.2, \quad
\lambda_{\text{grad}} = 0.2 \\
\mathcal{L}_{\text{SSIM}} &= 1 - \text{SSIM}(\text{pred}, \text{orig}) \\
\mathcal{L}_{\text{grad}} &= \frac{1}{2} \cdot 
\left\| \nabla_y \text{pred} - \nabla_y \text{orig} \right\|_1 + 
\frac{1}{2} \cdot \left(1 - \left\| \nabla_x \text{pred} \right\|_1 \right)
\end{aligned}
\end{equation}

Here, $\mathcal{L}_{\text{MSE}}$ is the mean squared error, $\mathcal{L}_{\text{SSIM}}$ controls the contribution of the SSIM (Structural Similarity Index Measure) loss , and $\mathcal{L}_{\text{grad}}$
weights the gradient loss for preserving edges and textures.

\vspace{1em}
\noindent The SSIM is computed as:
\noindent 
\begin{equation}
\begin{aligned}
\mathrm{SSIM}(x, y) &= \frac{(2\mu_x\mu_y + C_1)(2\sigma_{xy} + C_2)}{(\mu_x^2 + \mu_y^2 + C_1)(\sigma_x^2 + \sigma_y^2 + C_2)} \\
\mu_x &= \mathrm{mean}(x), \quad \mu_y = \mathrm{mean}(y) \\
\sigma_x^2 &= \mathrm{mean}\left((x - \mu_x)^2\right), \quad \sigma_y^2 = \mathrm{mean}\left((y - \mu_y)^2\right) \\
\sigma_{xy} &= \mathrm{mean}\left((x - \mu_x)(y - \mu_y)\right) \\
L &= \max(x, y) - \min(x, y) \quad \\
C_1 &= (0.01 \cdot L)^2, \quad C_2 = (0.03 \cdot L)^2 
\end{aligned}
\end{equation}

where $x$ and $y$ denote the predicted and original images, respectively.


\begin{table*}
\centering
\caption{Architecture of the U-Net model for image denoising.}
\label{tab:basic_unet}
\begin{tabular}{|c|c|c|c|}
\hline
\textbf{Module} & \textbf{Layer Type} & \textbf{Configuration} & \textbf{Output Shape} \\
\hline
\multicolumn{4}{|c|}{\textbf{Encoder (Downsampling path)}} \\
\hline
Down1 & Conv2D + SiLU & $1 \rightarrow 32$, kernel=5, padding=2 & $B \times 32 \times H \times W$ \\
Down2 & Conv2D + SiLU & $32 \rightarrow 64$, kernel=5, padding=2 & $B \times 64 \times H \times W$ \\
       & MaxPool2D & kernel=2, stride=2 & $B \times 64 \times \frac{H}{2} \times \frac{W}{2}$ \\
Down3 & Conv2D + SiLU & $64 \rightarrow 64$, kernel=5, padding=2 & $B \times 64 \times \frac{H}{2} \times \frac{W}{2}$ \\
       & MaxPool2D & kernel=2, stride=2 & $B \times 64 \times \frac{H}{4} \times \frac{W}{4}$ \\
\hline
\multicolumn{4}{|c|}{\textbf{Decoder (Upsampling path)}} \\
\hline
Up1 & Conv2D + SiLU & $64 \rightarrow 64$, kernel=5, padding=2 & $B \times 64 \times \frac{H}{4} \times \frac{W}{4}$ \\
     & Upsample (scale=2) + Skip Add & - & $B \times 64 \times \frac{H}{2} \times \frac{W}{2}$ \\
Up2 & Conv2D + SiLU & $64 \rightarrow 32$, kernel=5, padding=2 & $B \times 32 \times \frac{H}{2} \times \frac{W}{2}$ \\
     & Upsample (scale=2) + Skip Add & - & $B \times 32 \times H \times W$ \\
Up3 & Conv2D + SiLU & $32 \rightarrow 1$, kernel=5, padding=2 & $B \times 1 \times H \times W$ \\
\hline
\end{tabular}
\vspace{1mm}
\parbox{\textwidth}{\footnotesize Notes: Conv2D = 2D convolution layer for feature extraction; SiLU = activation function; MaxPool2D = downsampling by max pooling; Upsample = feature map upscaling; Skip Add = adding encoder features to decoder for detail recovery; B, H, W = batch size, height, width.}
\end{table*}

\begin{table*}
\centering
\caption{Architecture of the classifier model.}
\label{tab:combined_model}
\begin{tabular}{|c|c|c|c|}
\hline
\textbf{Module} & \textbf{Layer Type} & \textbf{Configuration} & \textbf{Output Shape} \\
\hline
\multicolumn{4}{|c|}{\textbf{SE-ResNet10\_1D Branch (for profile)}} \\
\hline
Init & Conv1D + BN & $1 \rightarrow 8$, kernel=7, stride=2, padding=3 & $B \times 8 \times L$ \\
     & AvgPool1D   & kernel=2, stride=1, padding=1 & $B \times 8 \times L$ \\
\hline
Block1 & Bottleneck1D + SE & $8 \rightarrow 16$, expansion=2 & $B \times 16 \times L$ \\
Block2 & Bottleneck1D + SE & $16 \rightarrow 16$, expansion=2 & $B \times 16 \times L$ \\
Block3 & Bottleneck1D + SE (downsample) & $16 \rightarrow 32$, expansion=2 & $B \times 32 \times L/2$ \\
Pool   & AvgPool1D   & kernel=2, stride=1, padding=1 & $B \times 32 \times L/2$ \\
\hline
Flatten & - & - & $B \times F_1$ \\
\hline
\multicolumn{4}{|c|}{\textbf{CNN\_Attention Branch (for time-phase)}} \\
\hline
Conv1 & Conv2D + BN & $1 \rightarrow 8$, kernel=(5,3), stride=1, padding=(2,1) & $B \times 8 \times H \times W$ \\
Pool1 & AvgPool2D & kernel=2 & $B \times 8 \times H/2 \times W/2$ \\
CA     & Channel Attention & reduction ratio=8 & $B \times 8 \times H/2 \times W/2$ \\
SA     & Spatial Attention & kernel=5 (along H) & $B \times 8 \times H/2 \times W/2$ \\
Conv2 & Conv2D + BN & $8 \rightarrow 16$, kernel=(5,3), stride=2 & $B \times 16 \times H/4 \times W/4$ \\
Pool2 & AvgPool2D & kernel=2 & $B \times 16 \times H/8 \times W/8$ \\
\hline
Flatten & - & - & $B \times F_2$ \\
\hline
\multicolumn{4}{|c|}{\textbf{Feature Fusion and Classifier}} \\
\hline
Fusion & Concatenate & $[F_1, F_2] \rightarrow 9824$ & $B \times 9824$ \\
FC1 & Linear + BN + ReLU + Dropout(0.2) & $9824 \rightarrow 512$ & $B \times 512$ \\
FC2 & Linear + BN + ReLU + Dropout(0.4) & $512 \rightarrow 128$ & $B \times 128$ \\
FC3 & Linear & $128 \rightarrow 2$ & $B \times 2$ \\
\hline
\end{tabular}
\parbox{\textwidth}{\footnotesize Notes: Conv1D / Conv2D = convolution layer for 1D / 2D inputs; BN = batch normalization; AvgPool1D / AvgPool2D = average pooling for downsampling; SE = squeeze-and-excitation attention; Bottleneck1D = Bottleneck block; Channel / Spatial Attention = attention mechanisms for channels / spatial dimensions; Flatten = reshape feature map to vector; Concatenate = combine features from different branches; Linear = fully connected layer; ReLU = activation function; Dropout = regularization by random feature masking; B, L, H, W = batch size, sequence length, height, width; F1, F2 = flattened feature dimensions from each branch.}
\end{table*}


\bsp	
\label{lastpage}
\end{document}